\documentclass[fleqn,usenatbib,useAMS]{mnras}

\usepackage{natbib}
\usepackage{newtxtext,newtxmath}

\usepackage[T1]{fontenc}
\usepackage{ae,aecompl}

\usepackage{graphicx}
\usepackage{amssymb,amsmath}
\usepackage{epstopdf}

\newcommand{\pd}[1]{\frac{\partial}{\partial #1}}
\newcommand{\fil}[1]{\langle #1 \rangle}

\title[Resolving the turbulent flow in the outskirts]{Adaptive mesh refinement simulations of a galaxy cluster merger -- I. Resolving and modelling the turbulent flow in the cluster outskirts} 

\author[L. Iapichino et al.]{L. Iapichino$^{1,2}$\thanks{E-mail:
    luigi.iapichino@lrz.de (LI)}, C. Federrath$^{3}$,  and R. S. Klessen$^{2,4}$ \\
\\
$^{1}$Leibniz-Rechenzentrum der Bayerischen Akademie der Wissenschaften, Boltzmannstr.~1, D-85748 Garching b.~M\"unchen, Germany\\
$^{2}$Universit\"at Heidelberg, Zentrum f\"ur Astronomie, 
Institut f\"ur Theoretische Astrophysik, Albert-Ueberle-Str.~2 and Philosophenweg 12, \\ D-69120 
Heidelberg, Germany\\
$^{3}$Research School of Astronomy and Astrophysics, Australian National University, Canberra, ACT 2611, Australia\\
$^{4}$Universit\"at Heidelberg, Interdisziplin\"ares Zentrum f\"ur Wissenschaftliches Rechnen (IWR), Im Neuenheimer Feld 205, D-69120 \\ Heidelberg, Germany 
}

\begin{document}
\label{firstpage}
\date{Accepted 2017 April 07. Received 2017 April 07; in original form 2016 May 13}
\pubyear{2017}

\pagerange{\pageref{firstpage}--\pageref{lastpage}} 

\maketitle

\begin{abstract}  
The outskirts of galaxy clusters are characterised by the interplay of gas accretion and dynamical evolution involving turbulence, shocks, magnetic fields and diffuse radio emission. The density and velocity structure of the gas in the outskirts provide an effective pressure support and affect all processes listed above. Therefore it is important to resolve and properly model the turbulent flow in these mildly overdense and relatively large cluster regions; this is a challenging task for hydrodynamical codes. 
In this work, grid-based simulations of a galaxy cluster are presented. The simulations are performed using adaptive mesh refinement (AMR) based on the regional variability of vorticity, and they include a subgrid scale model (SGS) for unresolved turbulence. The implemented AMR strategy is more effective in resolving the turbulent flow in the cluster outskirts than any previously used criterion based on overdensity. 
We study a cluster undergoing a major merger, which drives turbulence in the medium. The merger dominates the cluster energy budget out to a few virial radii from the centre. In these regions the shocked intra-cluster medium is resolved and the SGS turbulence is modelled, and compared with diagnostics on larger length scale.
The volume-filling factor of the flow with large vorticity is about $60\%$ at low redshift in the cluster outskirts, and thus smaller than in the cluster core. In the framework of modelling radio relics, this  point suggests that upstream flow inhomogeneities might affect pre-existing cosmic-ray population and magnetic fields, and the resulting radio emission.
\end{abstract}


\section{Introduction}
\label{intro}

The evolution of the large-scale structure of the universe proceeds through a sequence of merger events, by which collapsed objects contribute to form larger entities (e.g.~\citealt{o93,wbh93}). At the top of this hierarchy, clusters of galaxies reach a mass up to a few times $10^{15}$\,M$_{\sun}$. The baryonic gas falls in the corresponding potential wells and is shock-heated to temperatures up to about  $10^8$\,K in the intra-cluster medium (ICM). A fraction of the gas tracks the filaments interconnecting and surrounding the collapsed halos, and can be found in a phase of lower density and temperature than the ICM, called the warm-hot intergalactic medium (WHIM).

Since the hot ICM gas emits in the X-ray wavebands, mainly by thermal bremsstrahlung emission, and the X-ray emissivity scales with $n_{\text e}^{-2}$, where $n_{\text e}$ is the electron number density, the X-ray emitting gas in the outskirts is very hard to detect. It is therefore not surprising that observations of the cluster periphery  
have received attention only in recent years (see the review by \citealt{rbe13}).

The observations of the cluster outskirts have raised important questions about the physical conditions of the gas at those locations. In particular, the most striking discrepancy with analytical predictions \citep{tn01} and simulations (e.g., \citealt{vkb05}) is in the entropy profiles of the outskirts, flattening to values smaller than the theoretical expectations. In order to explain this feature, gas clumping has been invoked \citep{sam11,mnc13,iws14,wfs13,ves13,reb13,zck13}, as well as electron-ion equilibration timescales (see recently \citealt{anl14}, and references therein). An alternative possibility is that part of the pressure support in the cluster periphery comes from turbulent gas flows. This can be easily understood, since the outskirts are subject to accretion flows delivering material from the surrounding regions as part of cosmic structure formation, and had no time to settle in hydrostatic equilibrium, unlike the cluster core. This scenario is supported by many numerical simulations \citep{lkn09,vbk09,vbg11,pmq12,skn15}. Accretion  has  been  proposed  to  drive  turbulence  also  on  scales of  galaxies  or  molecular  clouds  within  these  galaxies  (see, e.g., the discussion by \citealt{kh10}). 

Probing the physics of the cluster peripheries with realistic numerical simulations has become mandatory for using galaxy clusters as tools in cosmology. Among the points touched above, the level of turbulence is of  special importance because the presence of non-thermal pressure will result in underestimating the cluster mass, if hydrostatic equilibrium in the ICM is assumed \citep{rtm04}. 
The first direct detection of turbulence-induced X-ray line broadening in the ICM, performed on the Perseus cluster by the {\it Hitomi} satellite \citep{hitomi16}, indicates that the correction due to turbulence pressure is not of leading order in the core regions of that relaxed object.  The unfortunate loss of the satellite leaves the problem of direct detection of turbulence in other clusters with different dynamical histories open, most likely until one of the next upcoming missions like {\it Athena}\footnote{Website: http://www.the-athena-x-ray-observatory.eu/} is launched. 

The important role of the turbulent kinetic energy component in the cluster outer regions has intriguing consequences for the study of non-thermal processes. Both MHD simulations (e.g., \citealt{dbl02,dt08,rkc08,ds09,xlc10,bds11,xlc11,mb15,bm15}) and analytical estimates\footnote{Also noteworthy in this framework is the development of laboratory experiments for the study of turbulence and turbulent magnetic amplification in astrophysical plasmas \citep{mdm14,mtb15}.} (\citealt{ib12}; see also \citealt{ssk13,sls13}) indicate that magnetic fields, amplified by the turbulent dynamo, can be in the range between 0.1 and 1\,$\mu$G. Indeed, observational analyses from different techniques \citep{ckb01,gmf06,bfm10} support this idea. Also, connected with shocks and turbulence is the acceleration of cosmic ray (CR) electrons \citep{bj14}. The presence of this particle component and of magnetic fields is indicated by the observations of diffuse emission in the radio wavelengths in a growing number of cluster peripheries; these objects are known as radio relics (\citealt{fgs08}, for a review). For some relics, X-ray observations have detected the presence of a shock corresponding to the location of the radio emission \citep{gvm08,fsn10,mmg11,atn12,ak13,bgb16,bse16}. Propagating shocks in the ICM have also an impact on the star formation rate of cluster galaxies \citep{ssr14}, which is in agreement with the dependence of star formation on Mach number in molecular clouds, studied by \citet{fk12} and \citet{rkb12,rbk14}. 

A leading model for relic emission implies the acceleration  of CR electrons at the shock (diffusive shock acceleration, henceforth DSA), and the subsequent synchrotron emission in the downstream magnetic field. 
The details of the acceleration and of the dynamics of CRs in relics go beyond the scope of the present study and are extensively referred elsewhere \citep{bbr12,krj12,bj14}. Here it is sufficient to recall that there has been tension between the DSA mechanism as formation model for the relics and observational data. For example, the acceleration efficiency of cluster merger shocks (with a Mach number of a few) has been questioned (e.g., \citealt{kr11}). An increasing interest \citep{pop13} is attracted by acceleration models invoking a pre-existing CR population, either produced by earlier stirring events or ejected by active galaxies \citep{smb15}. Turbulent reacceleration in relics has been proposed by \citet{fty15}. In principle, the propagation of a shock can thus expose the upstream conditions of the ICM, as far as CR population is concerned. Furthermore, we notice that inhomogeneities of the upstream magnetic field can mimic the same effect. 

Since shocks and turbulence in the ICM are related both to the acceleration of the CRs and to the amplification of the magnetic fields, we take a first step in the study of the problem and investigate the properties of the flow in the cluster outskirts purely from the viewpoint of hydrodynamics, putting aside the MHD treatment and the acceleration model.  This requires the choice of a numerical scheme able to properly model and resolve the ICM flow, and in particular in the outer cluster regions.

Dealing with the cluster outskirts and the surrounding filaments is extremely demanding for numerical simulations. 
Smoothed Particle Hydrodynamics is self-adaptive in simulations of collapsed objects like galaxy clusters, which is convenient to spatially resolve the dense central regions, but can be problematic in the mildly overdense surroundings (e.g.~\citealt{vdr11}). On the other hand, Adaptive Mesh Refinement (AMR) simulations can in principle allow to refine on any arbitrary variable, resulting in a matchless advantage in problems involving turbulent flows. 
Resolving turbulent flows has been motivating the AMR simulations performed by using refinement criteria based on gas velocity jumps \citep{vbk09} or on regional variability of vorticity modulus and compression rate (\citealt{in08,sfh09,ssi15}; see Section \ref{tools} for a detailed description). In recent years, the increased availability of computational power has revived the static grid approach (with some technical variants), in simulations able to provide an inertial range (the interval of length scales between the turbulence driving scale and the dissipation scale) of at least one decade in length scale, suitable to resolve the turbulent flow in the ICM and around the cluster \citep{vgb14,vbg14,m14,m15,vjb16}.

Previous numerical studies of turbulence in galaxy clusters \citep{isn11} point to an important difference between the ICM and the WHIM. In those simulations, the time evolution of the turbulence energy in the ICM shows a peak at redshift between $z \simeq 1$ and  $0.6$, while in the WHIM the turbulence energy grows steadily until the current epoch. The explanation for these different evolutions is provided by the turbulence injection mechanisms in the two gas phases: turbulent flow is induced mainly by cluster major mergers in the ICM, and by accretion through structure formation shocks in the WHIM. For the latter, and in particular for the cluster outer regions, a similar result has been developed analytically by \citet{clf11}.

The simulation discussed by \citet{isn11} makes use of a subgrid scale (henceforth SGS) model for the computation of the energy content on unresolved length scales \citep{snh06,mis09}. This tool, to be discussed in Section \ref{tools}, is very suitable in simulations of turbulent flows in contracting objects with a large dynamical range. On the other hand, a limit of that work is a modest effective spatial resolution ($48.8$\,kpc\,$h^{-1}$), necessary to simulate a relatively large cosmological volume with a feasible dynamical range. Zoomed simulations of single clusters, with a combination of static grids and AMR allowing a much finer spatial resolution, are thus needed for a more complete and complementary understanding of the cluster energy budget and its time evolution.

The current work and its companion study (henceforth Paper II) aim to focus on the cluster outskirts and further elaborate on the properties of the turbulence flow in those regions, as opposed to the cluster core, during a major merger. We perform a number of zoomed, grid-based cosmological simulations of the evolution of a massive cluster, including turbulence SGS model (\citealt{snh06}, with the addition of the improvements by \citealt{sf11}) and AMR criteria suitable for refining turbulent flows (Section \ref{tools}). By using these tools and analysing the flow properties, we will study the following relevant questions for the ICM physics and for the turbulence driving mechanisms during cosmological structure formation:
\begin{enumerate}
\item Is the turbulence in the cluster outer regions controlled by accretion of pristine gas, as \citet{isn11} stated for the WHIM, or is it rather affected by merger events occurring in the ICM? 
\item What are, during the cluster evolution, the dominant stirring modes (solenoidal or compressional)? 
\item How intense and volume-filling is the turbulent flow in the cluster outskirts?
\end{enumerate}
This first paper focuses mainly on the numerical methods, namely on the effectiveness of the AMR strategy based on regional variability of  vorticity in refining the cluster outer region, when compared with other methods based on overdensity (Section \ref{amr}). Furthermore, we will show that the employed turbulence SGS model provides a quantity (the subgrid turbulent kinetic energy, defined in Section \ref{sgs}) which is useful in the characterisation of the turbulent flow during the cluster evolution and, being computed locally (cell-by-cell), it is more easily accessible than global (volume-averaged) diagnostics. It will be discussed how these observables of turbulence on different length scales compare to each other during a merger event. We will limit the study of the general properties of turbulence in the ICM to the minimum which is needed to interpret the results of the SGS model, well aware that results on large scales are available in literature and cited above. In Paper II, more devoted to the physical interpretation of simulation results, we further analyse the velocity and density field both of the innermost and the outer cluster regions, to understand how the turbulent flow evolves during a major merger, and which stirring modes are dominant. 

The structure of the current paper is the following: in Section \ref{tools} we describe our numerical tools, in particular the AMR criteria and the turbulence SGS model. Our results are presented in Section \ref{results}, where we focus in particular on comparisons of different refinement strategies and diagnostics of the turbulent flow, both on large and subgrid scales.
The results are discussed in Section \ref{discussion}, and our conclusions are summarised in Section \ref{conclusions}.

\section{Numerical methods}
\label{tools}

The numerical simulations presented here have been performed using the {\sc enzo} code \citep{obb05,enzo14}, in a customised version based on the public release v.~2.1. {\sc enzo} is a hybrid, $N$-body plus hydrodynamical grid-based code featuring block-structured AMR \citep{bc89}. The $N$-body section of the code is based on a particle-mesh scheme \citep{he88}, and the hydrodynamics on the Piecewise Parabolic Method (PPM; \citealt{cw84}).

The simulated box has a size of $256$\,Mpc\,$h^{-1}$ on a side, and is evolved from redshift $z=60$ to $z=0$. The initial conditions have been produced according to the \citet{eh99} transfer function. We adopted the cosmological parameters coming from the {\it Planck} 2013 results \citep{planck14}, and consequently set $\Omega_\Lambda = 0.693$, $\Omega_{\text m} = 0.307$, $\Omega_{\text b} = 0.0482$, $h = 0.678$, $\sigma_8 = 0.826$, and $n=0.9608$.

Besides the turbulence SGS model (Section \ref{sgs}), we did not include any additional physics to our runs, as we are interested primarily in cluster regions and length scales where extensions beyond the "adiabatic" description of the fluid play a minor role (for some investigation, see \citealt{eve12}). We model the gas as a fluid described by an ideal equation of state, with $\gamma = 5/3$. 

The simulation setup has been optimised for resolving the formation of a single galaxy cluster in its cosmological environment. The simulation volume is resolved in $128^{3}$ grid cells and $128^{3}$ $N$-body particles modelling the dark matter (DM). Inside this root grid (whose resolution is also identified by the refinement level $l=0$) two additional static grids are nested around the box centre ($l=1$ and 2). Each of these nested grids is resolved in $128^{3}$ grid cells and $128^{3}$ $N$-body particles; they have a size of $128$\,Mpc\,$h^{-1}$ ($l=1$) and $64$\,Mpc\,$h^{-1}$ ($l=2$), respectively, with the $N$-body particle mass at $l=2$ being $8.5\times 10^9$\,M$_{\sun}$\,$h^{-1}$. In an innermost volume of $38.4$\,Mpc\,$h^{-1}$ on a side AMR is activated, according to the refinement strategies discussed below (Section \ref{amr}). The size of this region is chosen in such a way to be, at $z=2$, about four times larger than the Lagrangian volume of the $N$-body  particles that, at $z=0$, are part of the cluster. The finest allowed AMR level is $l=8$, corresponding to an effective spatial resolution of $7.8$\,kpc\,$h^{-1}$.

\subsection{Criteria for adaptive mesh refinement}
\label{amr}

In mesh-based codes AMR is a crucial tool for gaining the required dynamic range to model strongly clumped systems like galaxy clusters. However, by its very nature, AMR has issues in dealing with high refinement in extended regions of the computational volume. The choice of the refinement strategy is of great importance, also in order to avoid over-refinement on uninteresting locations: this would spoil the advantage of using AMR with respect to other alternatives (static grids, both monolithic or nested), and moreover would make the computation infeasible on modern computer architectures, where the amount of available memory per core is a significant constraint. 

A sensible and widely used AMR choice in simulations of cosmological structure formation is refinement on overdensity, referring both to the DM and the baryon component. This criterion is already implemented in {\sc enzo}, and a cell is flagged for refinement if the local density $\rho_\text{i}$, where "i" can indicate either baryons or DM, fulfils the following criterion: 
\begin{equation}
\label{threshold}
\rho_\text{i} > f_\text{i} \rho_0 \Omega_\text{i} N^l\,\,,
\end{equation}
where $\rho_0 = 3 H_0^2 / 8 \pi G$ is the critical density, $H_0$ is the Hubble parameter at the present epoch, $\Omega_\text{i}$ are the cosmological density parameters for either baryons or DM, and the refinement factor is $N=2$.
In this work the parameters $f_\text{i}$ for overdensity are set such that $f_{\text{b}} = f_{\text{DM}}$.

The choice of the ideal overdensity parameter depends on the kind of problem that one wants to address. The work of \citet{ons05} has shown that the DM refinement is critical for properly resolving substructures. Our choices of $f_\text{i}$ are between $2$ and $8$; we present them in detail in Table \ref{tab1sims}. 

Among the different choices for refinement criteria for intermittent turbulent flows \citep{knp06,vbk09}, the approach chosen in this work is based on the regional variability of so-called structural invariants of the flow, i.e.~on variables related to spatial derivatives of the velocity field. This method has been introduced by \citet{sfh09} and applied by \citet{ias08} and \citet{ssi15} in idealised subcluster simulations, and in full cosmological simulations by \citet{in08}, \citet{pim11}, and \citet{sab14}. Here for the first time it is optimised for the study of cluster outskirts.

According to this criterion, a cell is flagged for refinement if the local value of the variable $Q(\bmath{x},t)$ is 
\begin{equation}
\label{local}
Q(\bmath{x},t) \ge \langle Q \rangle_i(t) + \alpha \lambda_i(t)\,\,,
\end{equation}
where $\lambda_i$ is the maximum between the average $\langle Q\rangle $ and the standard deviation of $Q$ computed on a grid patch $i$, and $\alpha$ is a tunable parameter. The rationale of this selection procedure is somewhat similar to the method by \citet{zck12} for selecting density inhomogeneities, although in the present work we decide to use a different variable. Instead of density, we refine on the vorticity modulus, $Q =\omega^2$. Up to a factor of 2, this quantity is also known as enstrophy (e.g., \citealt{vjb16}). The vorticity is defined as the curl of the gas velocity:
\begin{equation}
\bmath{\omega} = \nabla \times \bmath{v} \, \,.
\label{omega}
\end{equation}
As for the refinement parameter, we empirically set the value of the threshold factor $\alpha = 6.5$ like in \citet{in08}, for an optimal compromise between computational feasibility and refinement efficiency.

Table \ref{tab1sims} presents an overview on the performed runs. All simulations were run from $z=60$ to $z=2$ using the refinement criterion based on overdensity with thresholds $f_\text{i} = 4$ (that is $OD4$), and then evolved further with the criterion specified in the Table. As one can see, the number of AMR cells at $z=0$ (third column) spans almost one order of magnitude for the simulation sample. The fourth column reports $f_\text{V}$, defined as the fraction of the computational volume which is resolved at the highest refinement level $l=8$ (normalised to the volume where AMR is allowed). This quantity (anticipated here from Fig.~\ref{amr-volume}) has also a wide range of values within our sample, and together with the previous variable can be used to estimate  the computational cost of the simulations presented here.     

\begin{table}
\caption{Summary of the four simulations analysed in the present work. All runs follow the evolution of the same realisation of a cosmological volume and differ only for the AMR criterion from $z=2$. The AMR overdensity factor $f_\text{i}$, both for DM and baryons (third column), used in the runs is indicated for clarity also by the number in the simulation name. The turbulence SGS model is used in all simulations.}
\centering
\begin{tabular}{ccccc}
\hline
Simulation & $f_\text{i}$ & Cells [$\times 10^6$]   & $f_\text{V}$ [$\times 10^{-7}$] & AMR \\
 & & with $l \geq 3$ & ($l=8$) & criteria \\ 
\hline 
$OD4$+  & 4 & 13.35 & 148.9 & OD and \\
               &   &       &       &  vorticity\\
$OD2$ & 2 & 8.34 & 28.2 & OD only \\
$OD4$ & 4 & 3.69 & 6.23 & OD only \\
$OD8$ & 8 & 1.76 & 0.76 & OD only  \\
\label{tab1sims}
\end{tabular}
\end{table}

\subsection{Modelling of unresolved turbulence}
\label{sgs}

Our hydrodynamical code makes use of a SGS model for unresolved turbulence. This  tool has been described first by \citet{snh06} as an essential part of the flame propagation model in simulations of Type Ia supernovae \citep{snhr06}. In simulations of the evolution of the cosmic large-scale structure it has been used among others by \citet{mis09},  \citet{isn11,ivb13}, \citet{bsn14}, and \citet{sab14}. We notice that the adoption of turbulence SGS models has become more frequent in computational astrophysics in recent years (see e.g.~\citealt{sb08,cph13}), with pioneering work in SPH \citep{sws10}.

While the details of the model are contained in many of the works cited above (and we additionally refer the reader to \citealt{s15} for a recent review), we recall here only the concepts  that are crucial for a thorough understanding of what follows in this paper.

The turbulence SGS model is based on the decomposition of the fluid equations governing a physical system into a large-scale and a small-scale (unresolved) part, as described by the \citet{Germano1992} formalism. This decomposition, when kept implicit in mesh-based simulations, consists merely in neglecting the effect of unresolved scales, which is equivalent to claiming  that the numerical viscosity can mimic its effect. A more elaborate approach, however, is to explicitly select a filtering length scale and then to decompose any density-weighted variable of interest $f$ into a smoothed (i.e.~large-scale) part $\fil{f}$ and a fluctuating part $f'$, varying only on length scales larger than the filter \citep{FAVRE1969}. In this way, a filtered variable $\hat{f}$ is defined by the relation:
\begin{equation}
\fil{\rho f} = \fil{\rho} \hat{f} \Rightarrow \hat{f}=\frac{\fil{\rho f}}{\fil{\rho}}\,\,.
\label{filter}
\end{equation}
When applied to the equation of fluid dynamics for a compressible, viscous, self-gravitating fluid, this formalism leads to new terms in the momentum and energy equations (see \citealt{s15} for details), describing the interaction between resolved and unresolved scales. In particular, the filtered kinetic energy is expressed by 
\begin{equation}
\hat{e}_{\text{kin}} = \frac{1}{2}\hat{v_i}\hat{v_i} +  \frac{1}{2}\hat{\tau}(v_i,v_j)/ \fil{\rho} \, \,,
\label{ekin}
\end{equation}
where the first contribution on the right-hand side is the resolved kinetic energy. The second term contains the second-order moment of the velocity field $\hat{\tau}(v_i,v_j)=\fil{\rho v_i v_j} - \fil{\rho} \hat{v_i} \hat{v_j}$, and the trace of $\hat{\tau}(v_i,v_j)/ \fil{\rho}$ can be interpreted, like in \citet{Germano1992}, as the square of the SGS turbulence velocity $q$, so that 
\begin{equation}
e_{\text{t}}=\frac{1}{2}q^{2}=\frac{1}{2}\hat{\tau}(v_i,v_i)/\fil{\rho}
\end{equation}
defines the SGS turbulence energy. Following from the definition of $e_\text{t}$, and given that the trace of $\hat{\tau}(v_i,v_j)$ is added to the thermal pressure in the filtered momentum equation \citep{s15}, one can define the SGS turbulence pressure, associated with the unresolved turbulent velocity fluctuations on length scales smaller than the filter as
\begin{equation}
p_{\text{t}}=\frac{2}{3}\fil{\rho}e_{\text{t}}.
\label{pturb}
\end{equation}

The SGS turbulence energy is governed by the additional fluid equation,
\begin{equation}
\pd{t}\fil{\rho}e_{\text{t}}+\pd{r_j}\hat{v}_j
\fil{\rho}e_{\text{t}}=\ \mathcal{D}+\Sigma+\Gamma-\fil{\rho}(\lambda+\epsilon)\,\,,
\label{eq:etsum}
\end{equation}
where the terms $\mathcal{D}$, $\Sigma$, $\Gamma$, $\lambda$ and $\epsilon$ at the right-hand side are source and transport terms related to the scale separation. We refer the reader to \citet{mis09} for their definition and description. These terms are expressed by empirical closures, and their formulation represents the SGS model. The closures used in this work are the same as in \citet{mis09} and \citet{isn11}, with the exception of the formulation for the SGS turbulence stress tensor (contained in the definition of $\Sigma$, according to the so-called eddy viscosity closure), which follow the improved prescription by \citet{sf11}.
 
An essential feature of the turbulence SGS model in its implementation within an AMR code is the consistent bookkeeping of the energy components, in particular of the energy exchange between the resolved kinetic energy and the turbulence SGS energy due to a variation of the filtering length scale, at grid refinement or derefinement. Only at that step of the algorithm, one needs to resort to the additional assumption of Kolmogorov scaling of the turbulent energy \citep{k41}, according to the procedure described in detail by \citet{mis09} and \citet{isn11}. As Kolmogorov scaling is appropriate for incompressible turbulence, this choice is justified on the length scales of the order of several kiloparsec, near the spatial resolution of our simulations, whereas turbulence is nearly transonic up to slightly supersonic on injection length scales (see later in Section \ref{main-result}).

\section{Results}
\label{results}

\subsection{Properties of the simulated cluster}
\label{features}

With the use of preliminary, low-resolution simulations, the initial conditions of the runs have been spatially shifted (making use of the periodic boundary conditions) in such a way to form the most massive structure at the box centre, with the nested static grids and the AMR region enclosing this location. This cluster, at $z=0$ and in run $OD4$+, has a virial mass $M_\text{vir} = 0.95\times 10^{15}$\,M$_{\sun}\,h^{-1}$ and a virial radius $R_\text{vir} = 2.13$\,Mpc\,$h^{-1}$; these values differ for the other runs at a level of only a few per cent.
Throughout this work we will keep a definition of cluster virial radius by setting, like e.g.~in \citet{bn98}, the virial overdensity as $\Delta^c_\text{vir}(z = 0) \approx 102$, where the overdensity $\Delta$ is defined as the ratio between density and $\rho_c(z)$, the critical density of the Universe at redshift $z$. For sake of comparison with previous studies, $R_\text{vir} \approx 1.36\,R_{200}$ \citep{rbe13}, where the radius $R_{\Delta}$ at overdensity $\Delta$ is defined by the equation 
\begin{equation}
\frac{M(<R_{\Delta})}{4/3\ \pi R_{\Delta}^3} = \Delta \rho_c(z)\, , 
\label{eq:vir}
\end{equation}
here $M(<R_{\Delta})$ is the cluster mass enclosed within a sphere of radius $R_{\Delta}$.

\begin{figure*}
\includegraphics[width=\linewidth]{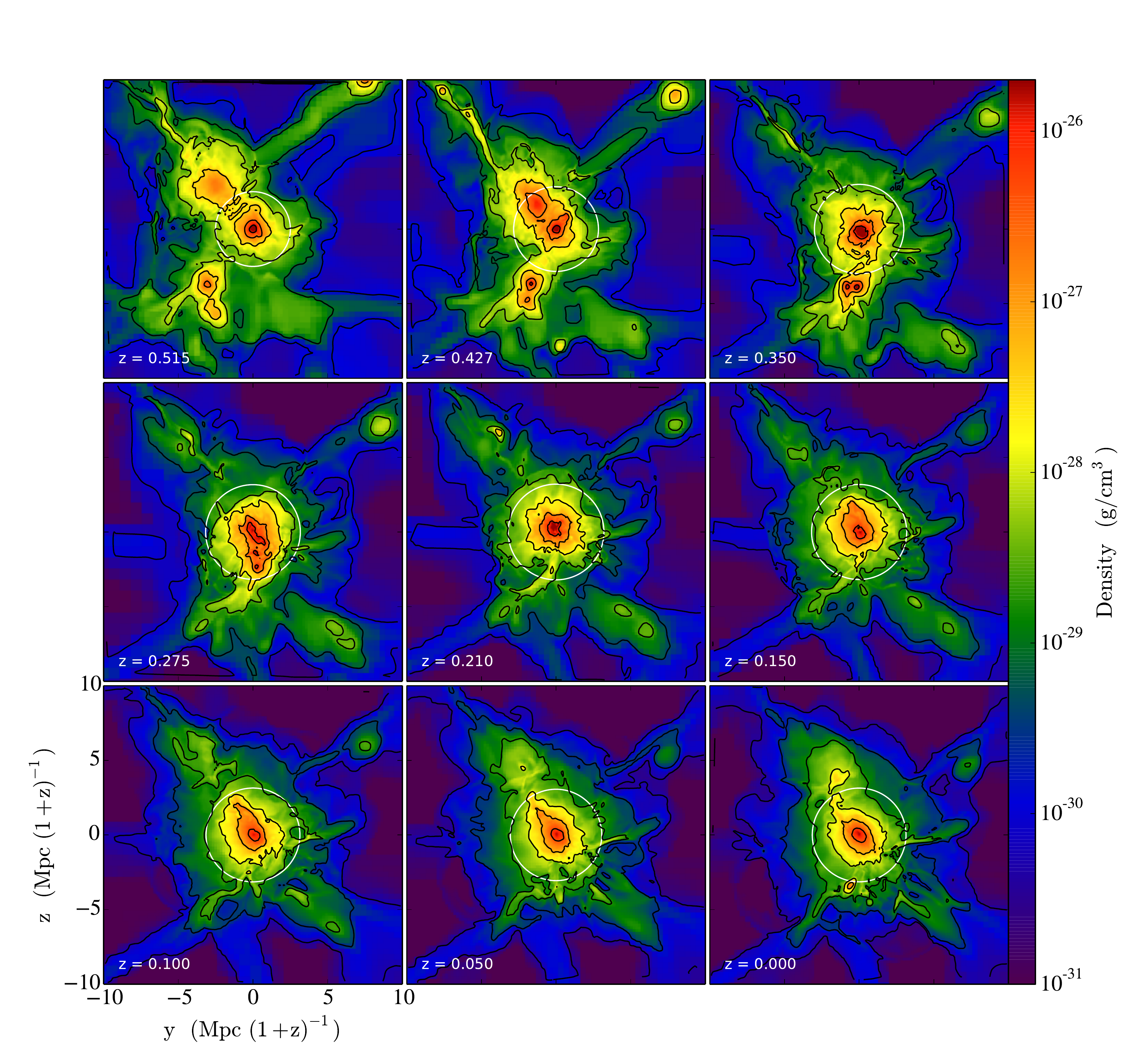}
\caption{Time evolution of the main cluster formed in run $OD4$+, as a series of slices showing gas density. Each slice is perpendicular to the $x$-axis, centred on the cluster centre as given by the {\sc hop} halo finder, and with a size of $(20$\,Mpc\,$h^{-1})^2$ comoving. In each panel the redshift $z$ is indicated, the white circle encloses the virial radius, and the black contours are density contours.}
\label{dens-evolution}
\end{figure*}

The analysis of the mass accretion history of this cluster, performed with the {\sc hop} algorithm \citep{eh98}, clearly shows the cluster undergoes a major merger between $z=0.515$ and $0.427$, with a mass ratio close to 1. The merger plane is almost perpendicular to the $x$-axis of the computational box, therefore in Figure \ref{dens-evolution} 
the merger is optimally shown by slices on the $yz$-plane. In this time sequence of gas density slices the two subclusters can be still distinguished at $z=0.427$. 

\begin{figure*}
\includegraphics[width=\linewidth]{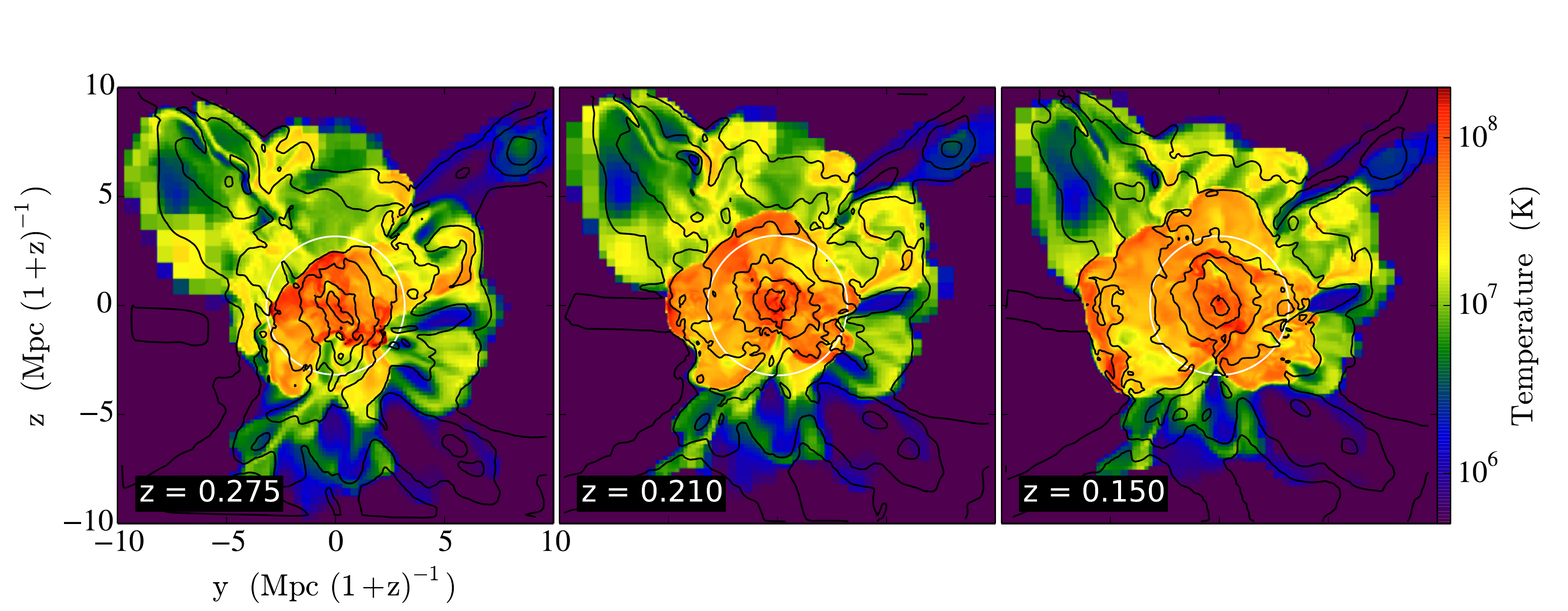}
\caption{Like in the central row of Figure \ref{dens-evolution}, but showing gas temperature, with values given by the colour bar.}
\label{temp-evolution}
\end{figure*}

Figure \ref{temp-evolution} shows a better view of the merger event through temperature slices. In those panels, the launched merger shock is located ahead of the high-temperature, shock-heated region, propagating outside. Around $z=0.210$, the shock crosses the virial radius, then propagates into the outskirts, where it remains visible until $z=0$. During its propagation inside the virial radius the merger shock has a Mach number initially ($z=0.35$) between $2.5$ and $4$, and at later times ($z= 0.275$) between $4$ and $6$, the range of values depending on the different shock regions. 
Interestingly, a visual inspection of Figure \ref{dens-evolution}  shows that, after the major merger, the cluster accretes smaller clumps (approaching from below until $z=0.275$). These multiple minor mergers lead to additional mass growth of $1.4\times 10^{14}$\,M$_{\sun}\,h^{-1}$ between $z=0.350$ and $0.275$. This process launches a second shock, weaker than the previous one and hardly visible inside the virial radius in Figure \ref{temp-evolution} at $z=0.15$, with a Mach number $\mathcal{M} \lesssim 3$. Such a complex merger scenario is ideal for studying the stirring of turbulence in the ICM and somewhat complementary to the works of \citet{in08} and \citet{mis09}, based on the analysis of a relaxed galaxy cluster.

\subsection{Comparison of mesh refinement strategies}
\label{comparisons}

\begin{figure}
  \includegraphics[width=\columnwidth]{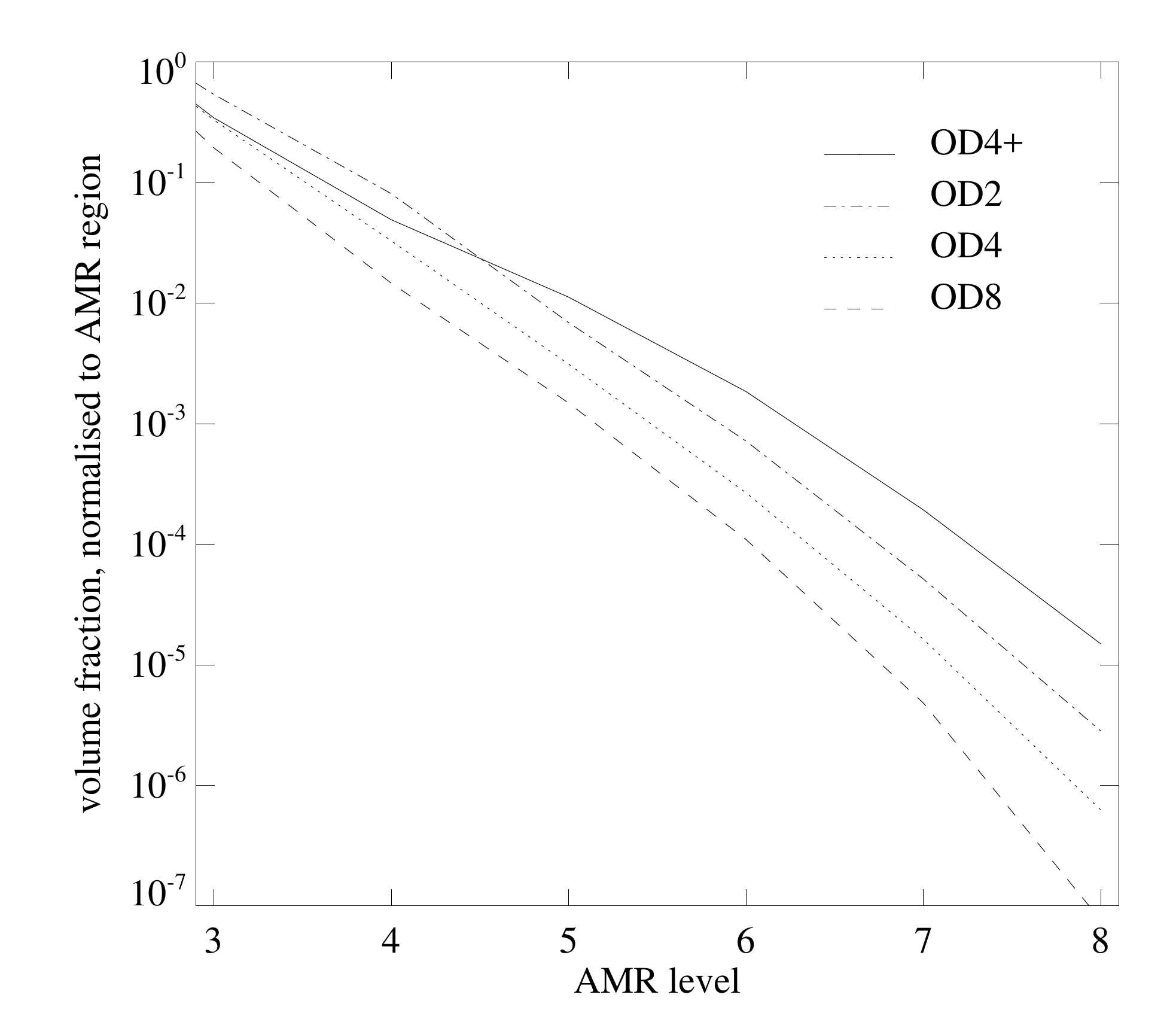}
  \caption{Volume fraction of the region where AMR refines cells at level $l \geq n$ ($n$ in the abscissa), for the different simulations (indicated in the legend) at redshift $z=0$.}
  \label{amr-volume}
\end{figure}

The simulations presented in this work differ from each other only in their refinement strategy and resulting ability in resolving the formation of cosmological structure. A first insight on the different AMR effectiveness has been provided by the number of computational cells in Table \ref{tab1sims}. 
Besides the total number of grid cells at $z=0$, an interesting indicator of the refinement effectiveness is the volume fraction refined at an AMR level $l \geq n$ (Figure \ref{amr-volume}). This is normalised to the volume of the part of the computational domain were AMR is allowed (a cube with size of $38.4$\,Mpc\,$h^{-1}$), so that this volume is equal to unity for $l=2$. One can see that the AMR criterion in run $OD8$ has the typical performance of simulations of cosmological large-scale structure, with every level occupying a volume approximately one order of magnitude smaller than its coarser one. Smaller overdensity thresholds result in refinement of larger volumes: the run $OD2$ refines a volume that is a factor of 37 larger than in run $OD8$ at the finest level $l=8$. The refinement on  vorticity is effective especially on large AMR levels: the run $OD4$+  refines more than $OD2$ from $l>5$, and provides an additional factor of 5.3 more resolved volume at $l=8$ than the latter. 

\begin{figure}
  \includegraphics[width=\columnwidth]{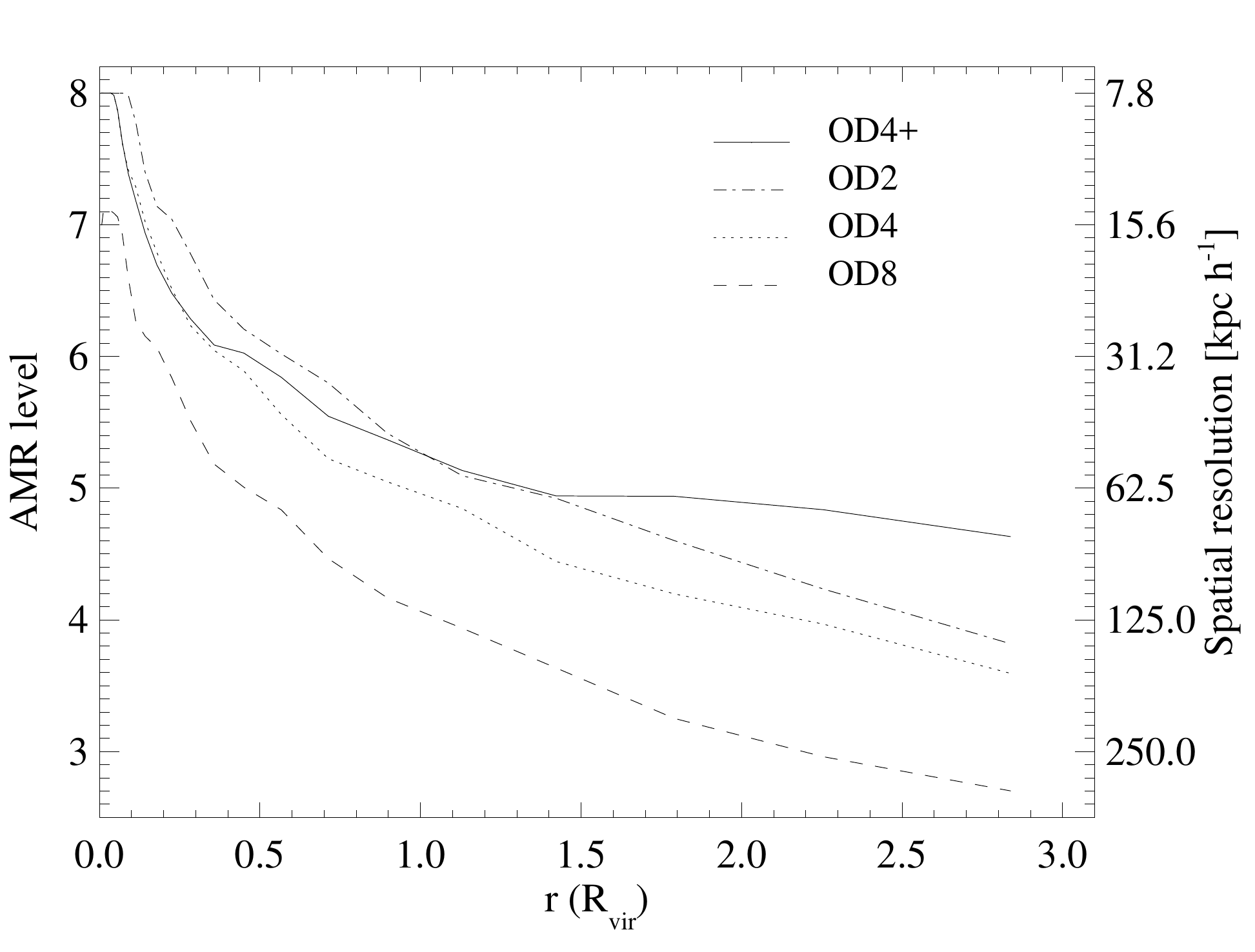}
  \caption{Radial profiles, around the centre of the cluster described in Section \ref{features}, of the volume-weighted AMR level at redshift $z=0$, for the simulations indicated in the legend. The scales on the $y$-axis show the AMR level (left-hand side) and the corresponding spatial resolution (right-hand) side.}
  \label{amr-profile}
\end{figure}

We emphasise that this strategy is especially successful in the refinement of the cluster outskirts. In order to demonstrate this, the cluster radial profiles of the mass-weighted AMR level at $z=0$ are shown in Figure \ref{amr-profile}. The profiles show how standard refinement criteria based on overdensity degrade their performance with distance from the cluster centre, and that even very permissive thresholds like in $OD2$ are of relatively little help outside $R_\text{vir}$. The refinement level in run $OD4$+  is smaller than that of $OD2$ at $r<0.6\ R_\text{vir}$, comparable for $0.6 < \ r/R_\text{vir} < 0.9$, and better for $ r > 0.9\ R_\text{vir}$. Not only the value of the AMR level is larger for the run $OD4$+, but we notice also that the slopes of the radial profiles for the other criteria are so low that only an unfeasibly small overdensity threshold (if any at all) would match the resolution of run $OD4$+ in the outskirts.
At $r = 2\ R_\text{vir}$ the refinement level of run $OD4$+  is larger by $\Delta l = 1.4$ and $0.8$ compared to runs $OD8$ and $OD2$, respectively, corresponding to a better resolution by a factor of $2^{\Delta l} = 2.64$ and $1.74$ for the two cases. In the best-resolved run, the mass-weighted AMR level within $r = 2\,R_\text{vir}$ is better than $l=5$, corresponding to a spatial resolution of $62.5$\,kpc\,$h^{-1}$.

\subsection{Resolving turbulence stirring in the cluster outskirts}
\label{filament}

\begin{figure*}
\includegraphics[width=0.97\linewidth]{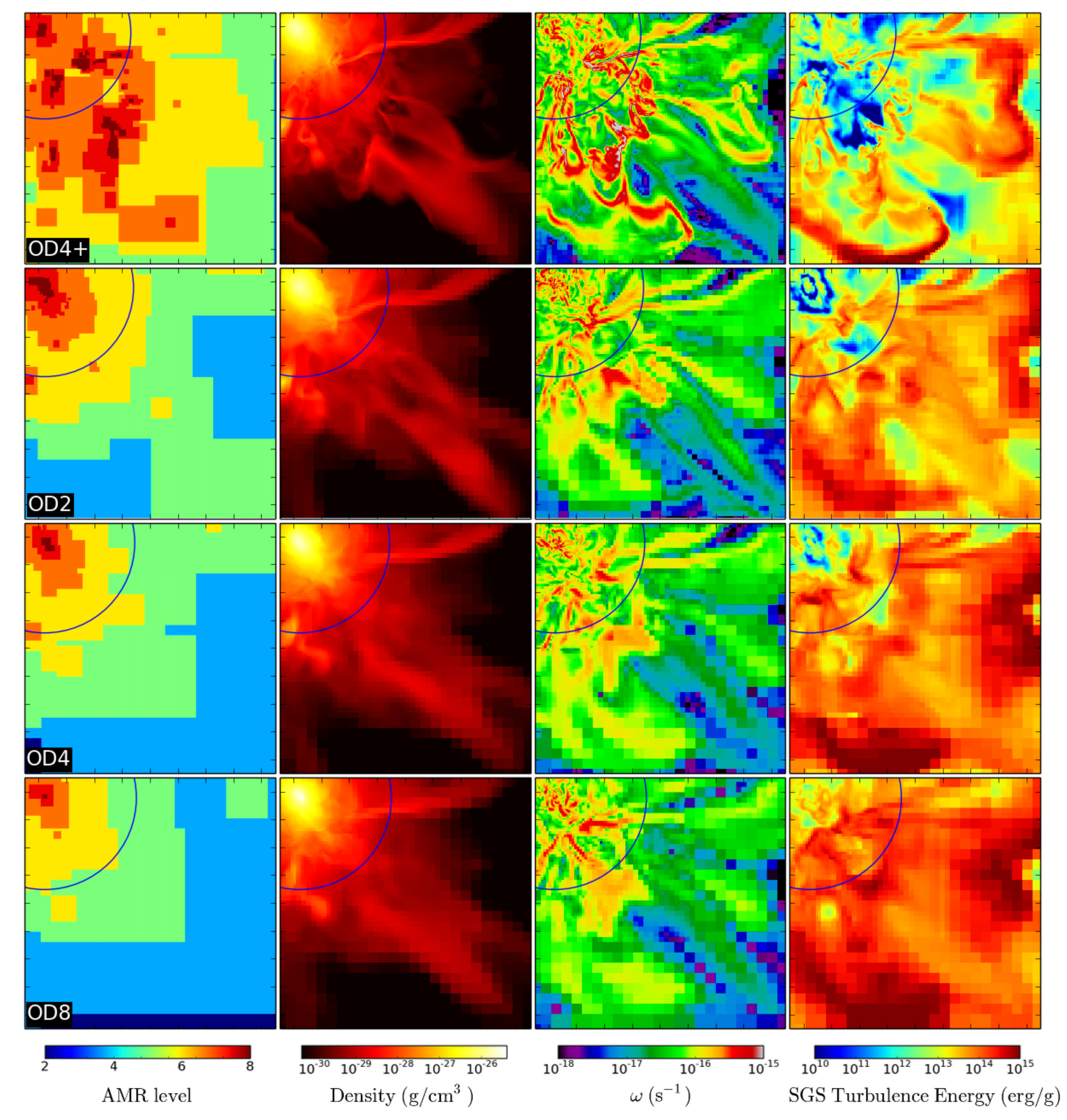}
\caption{From left to right: slices at $z=0$ of AMR level, gas density, vorticity modulus and SGS turbulence energy, for runs $OD4$+, $OD2$, $OD4$, and $OD8$ (top to bottom). The slices have a size of $9$\,Mpc\,$h^{-1}$ on a side. The blue circle indicates the location of the virial radius (and help locate the cluster centre).}
\label{slices-refin}
\end{figure*}

The main objective of AMR is to improve the resolution of the simulated physical system through a finer grid meshing. On the other hand, it is difficult to determine how effective an AMR strategy is in correctly capturing a turbulent flow, when simulating complex systems like galaxy clusters. Since the scope of the present work is to resolve turbulent flows, it is important that the proposed AMR strategy results in a good spatial coverage of the stirring agents, not only in the ICM but also in the cluster outskirts. The potential of the AMR based on regional variability of vorticity induced by subcluster motion has been shown by \citet{ias08} and \citet{ssi15} in idealised simulations, and by \citet{in08} in full cosmological simulations. Moreover, stirring driven by propagating merger shocks has been addressed by \citet{pim11}. Rather than repeating these analyses, here we provide a different example, and focus on the resolution of turbulence driven by infalling cosmic filaments onto the ICM. 

Cosmic filaments interconnect galaxy groups and clusters and contribute significantly to their mass accretion. Their role as turbulence drivers has been recently emphasised by \citet{zdb15}, and put into relation with other properties of galaxy clusters like their classification as relaxed objects or the presence of a cool core. At the interface where filaments impinge on the ICM the gas is shocked, further contributing to thermalisation and flow stirring. Moreover, the interaction between propagating shocks and filaments drives hydrodynamical shearing instabilities, which have been related to radio-bright notches at the edges of some observed radio relics \citep{pim11}. 

In Figure \ref{slices-refin} one can see, in the density slices, several examples of filaments, like the network located from the lower right corner to the upper left, or the thin structure coming from the upper right corner. Because these filaments are only moderately overdense, their spatial resolution depends crucially on the employed refinement criterion. The AMR used in run $OD4$+  outperforms the other strategies, as clearly visualised in the AMR level slices on the left-hand side, 
not only in terms of the morphology of the filaments (density slices, second column of Figure \ref{slices-refin}) but also for the small-scale vorticity (third column of Figure \ref{slices-refin}). A filamentary pattern typical of a turbulent flow extends well beyond the virial radius (up to about $2\,R_\text{vir}$ is visualised in the slices, and analysed in the following). Idealised simulations of supersonic turbulence show very similar patterns of enhanced vorticity at the locations of shocks (see e.g., figure 2 in \citealt{frk10}). We can therefore conclude that the refinement criterion based on vorticity is adequate for resolving turbulent driving.

In the last two columns of Figure \ref{slices-refin}, two different cell-based diagnostics of the turbulent flow are compared, namely the vorticity modulus of the flow and the SGS turbulence energy. The first one is a local indicator of spatial fluctuations of velocity, typical of turbulence, and is therefore a widely-used quantity for the study of the properties of flow, either in its basic definition or by isolating single terms of it (e.g.~\citealt{m14,vjb16}). A somewhat similar behaviour can be seen in the evolution of the SGS turbulence energy $e_\text{t}$, which is computed cell-wise by the turbulence SGS model employed in this work. These two quantities are indeed linked, as showed in \citet{isn11}  by the correlation of the terms expressing the resolved and SGS turbulent pressure (their figure 9). The most important conceptual difference between $\omega$ and $e_\text{t}$ is the dependence of the latter on grid resolution: in its definition,  the cell size acts as the filtering length scale introduced  in Section \ref{sgs}. The key consequence of this definition is that, in multi-resolution and/or AMR grids, cells at different resolution convey different information (i.e referring to different cutoff scales) about the subgrid scale turbulence. The meaningful use of this variable involves therefore some additional step for correctly interpreting the results. This may seem cumbersome at first, but we will show (Section \ref{shells}) that the analysis of $e_\text{t}$ provides similar results to the ones coming from other flow diagnostics, with the advantage of being computed locally on the grid without requiring data post-processing. The reason for it is that the properties of large-scale turbulence are imprinted onto the smallest scales through the turbulence cascade.  

In considering the slices of $e_\text{t}$ in Figure \ref{slices-refin}, we notice that the SGS turbulence energy tracks well the large-scale structure surrounding the cluster, in particular some filaments down to within the virial radius. This happens because $e_\text{t}$ correlates with shocks, in particular with the merger shock and the external ones, surrounding the cluster and the filaments. It is expected to find a larger $e_\text{t}$ in the post-shock regions \citep{pim11, ib12}, because shocks inject vorticity through the baroclinic mechanism (e.g., \citealt{rkc08,vjb16}). Moreover, it has been demonstrated that the turbulence SGS model with the closure prescription by \citet{sf11}, implemented in the simulation code used for this work, is reliable in computing the SGS energy in compressible flows, like the ones in the vicinity of shocks. 

To complete the comparisons performed in this Section, the last column of Figure \ref{slices-refin} presents an interesting consequence of the changing resolution level across the simulations, namely the variation in the SGS turbulence energy. Not surprisingly, in general $e_\text{t}$ is smaller in run $OD4$+, where the resolution is the highest, and therefore the unresolved part of the turbulent cascade is the smallest with respect to the other runs. For the same reason, the relatively large values in the outskirts of run $OD8$ mean that the flow in those regions is severely underresolved. These effects must be kept in mind for a correct interpretation of the results of the turbulence SGS model, as will be discussed in the following.

\subsection{Evolution of the kinetic energy on resolved and subgrid scales}
\label{shells}

In the previous Section the SGS turbulence energy $e_\text{t}$ has been compared with another indicator of the flow, the vorticity modulus. Here we compare $e_\text{t}$ and other diagnostics of the gas energy content, by taking averages over control volumes, and studying the different ways they evolve during a major merger.

The analysis presented in this Section makes use of a somewhat arbitrary distinction between the core region of the galaxy cluster and its outskirts. This definition is based on the cluster centre (defined as the location of highest DM density) and virial radius at every data output: the cluster "core" is enclosed in a sphere with radius of $0.5\,R_\text{vir}$ around the cluster centre, while the outskirts are defined by the spherical shell ranging from $0.5$ to $2\,R_\text{vir}$. This definition can be problematic during the merger phases where spherical symmetry is not a good approximation. However, tests with different values of boundary radii showed that the results are not significantly affected.

As for the quantities included in the comparison, besides $e_\text{t}$ we will make use of the specific internal energy $e_\text{int}$, and of the kinetic energy $e_\text{kin}$. The internal energy is a useful comparison term in our analysis, because in virialised objects like galaxy clusters there is a strong link between development of bulk flow and turbulence and gas thermalisation during mergers. In getting a definition of kinetic energy, one should keep in mind that the velocity of the centre of mass of a cluster in a cosmological box can be of several hundred km\,s$^{-1}$, which is of the same order of magnitude as the typical flow velocities. To get a cleaner indicator, the centre of mass velocity (computed by the {\sc hop} halo finder in the {\tt yt} toolkit) is subtracted from the velocity components, before computing the specific kinetic energy.

\begin{figure}
  \includegraphics[width=\columnwidth]{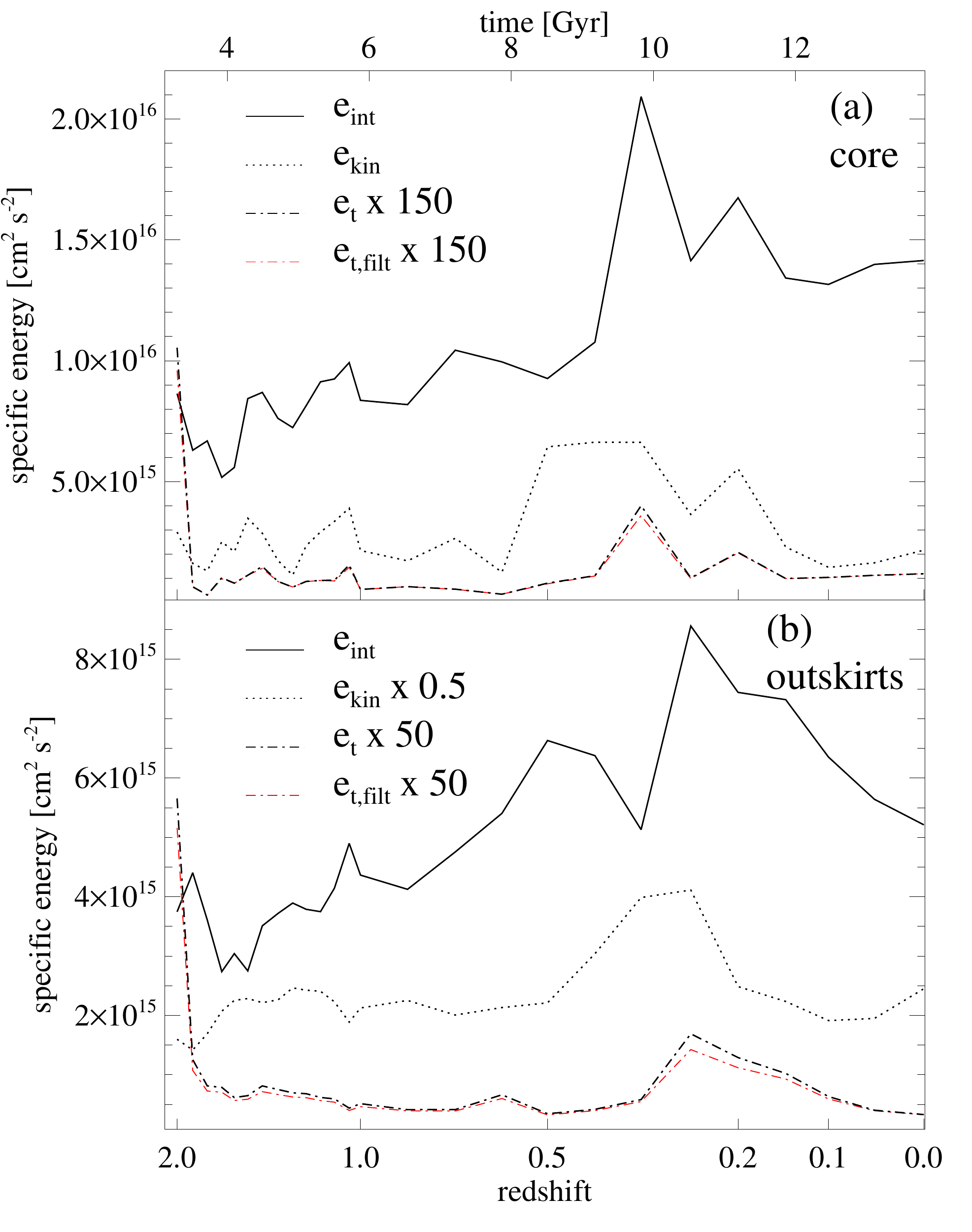}
  \caption{Top panel $(a)$: time evolution in the cluster core of the energy components (defined in the main text) $e_\text{int}$ (solid line), $e_\text{kin}$ (dotted line), $e_\text{t}$ (dot-dashed line) and $e_\text{t,filt}$ (dot-dashed red line), in the run $OD4$+. As indicated in the legend, some lines are scaled by arbitrary numerical factors for a better readability of the plot. Bottom panel $(b)$: same  as upper one, but for the cluster outskirts. }
  \label{energy-shells}
\end{figure}

We consider the evolution of energy in the core (Figure \ref{energy-shells}$a$) first. From the time evolution of $e_\text{int}$ one can recognise the merger sequence described in Section \ref{features}. After early fluctuations due to the initial cluster growth, the internal energy in the cluster core shows in fact a late double peak, related to the propagation of merger shocks released by the major and second merger, respectively. The double peak is also  apparent in the time evolution of the  SGS turbulence energy.
The steep decrease of $e_\text{t}$ right after $z =2$ is purely numerical, and caused by the activation of the AMR based on  vorticity at that time. Different from the previous two quantities, the kinetic energy $e_\text{kin}$ 
shows an increase that starts earlier than the double peak (already at $z = 0.5$). We will infer later that this is caused, at least in its first part, by the flow of filament gas and subclumps at merger stage, rather than by the merger shock propagation itself. 

The effect of the major merger is then visible in the outskirts (Figure \ref{energy-shells}$b$) with some delay, due to the propagation of the merger shock outwards; the second minor merger has nearly no impact outside the cluster core. Also here, the peaks of $e_\text{int}$ and $e_\text{t}$ at $z=0.275$ coincide, but the peak of the kinetic energy is broader. From this analysis we can draw the conclusion that the kinetic energy is sensitive both to the stirring caused by the cluster merger and by the subsequent merger shock propagation. The SGS turbulence energy, being the quantity most sensitive to smallest length scales, seems affected mostly by turbulence injected at shocks.
Since shocks also convert kinetic into internal energy, the correlation between $e_\text{t}$ and $e_\text{int}$ is readily explained. 

It has been verified that the correlation between shocks and turbulence injection is not biased by numerical effects. In Figure \ref{energy-shells} we also show the time evolution of $e_\text{t,filt}$, computed from $e_\text{t}$ by filtering out the cells belonging to the numerically smeared shocks (detected with the finder from \citealt{soh08}). The evolution of $e_\text{t,filt}$, both in the core and outskirts, is basically identical to that of $e_\text{t}$, and the same is true for the other energy components (not shown here). This result does not come unexpected, because the performance of the turbulence SGS model at shocks in compressive flows has been extensively tested by \citet{sf11}.

\subsection{Turbulent velocities on different scales}
\label{velocities}

The numerical values from Figure \ref{energy-shells} are better readable, and more directly comparable with observational data and predictions for cluster turbulence, when they are expressed in terms of velocity, rather than energy. This is done in Table \ref{turbvel}, where the velocities computed from the mass-weighted values of the energies are reported. We select the values at the redshift with largest $e_\text{t}$ and at $z=0$, as representative of a time with actively stirred and one with decaying turbulence, respectively. The averages of $q$ have values which are at least an order of magnitude lower than those of $v$. 
This is expected, because the velocity fluctuations probed by $q$ are on much smaller scales than those probed by $v$. 
Indeed, the ratio $R_{\text l}$ of length scales between the turbulence injection length scale 
and the effective spatial resolution of the simulation is approximately of the order of $100$. Assuming a velocity scaling $v \sim l^n$, a velocity ratio  
\begin{equation}
v/q \sim R_{\text l}^n
\label{eq:velratio}
\end{equation}
is expected, with $n=1/3$ or $1/2$ for the incompressible, Kolmogorov or compressible, Burgers turbulence, respectively (\citealt{k41,f95,f13}; cf.~also Section \ref{turb}). Given the flow properties in the cluster core and the outskirts, the two cases can be seen as extremes in our problem, with the former more relevant to the smaller length scales. For these two cases, Equation (\ref {eq:velratio}) provides $v/q \simeq 5$ and $10$, respectively, to be compared with a value around $15$ (see Table~\ref{turbvel}). The agreement with the simulation data within a factor of $2$ to $3$ is satisfactory, considering that this just as a rough estimate, and that large-scale velocities have a component of laminar motions in addition to the turbulent one.

\begin{table}
\caption{Velocities computed from the mass-weighted energy data in Figure \ref{energy-shells}, for the run $OD4$+. The reported velocities are $v$ (computed from  $e_\text{kin}$) 
and $q$ (from  $e_\text{t}$). The velocities are reported at $z=0$ and for the redshift of maximum  $e_\text{t}$ in the core/outskirts.}
\centering
\begin{tabular}{lcc}
\hline
 & $v$ &  $q$ \\ 
 & $[$km\,s$^{-1}]$  & $[$km\,s$^{-1}]$ \\ 
\hline
core & &  \\
$z=0$ & 658   & 39.7 \\ 
$z=0.350 $  & 1151    & 73.1 \\ 
\hline 
outskirts &  & \\
$z=0$ & 990    & 36.3 \\ 
$z=0.275 $  & 1283   & 82.2 \\ 
\label{turbvel}
\end{tabular}
\end{table}

\subsection{Evolution of vorticity and volume filling factor}
\label{vorticity}

\begin{figure}
  \includegraphics[width=\columnwidth]{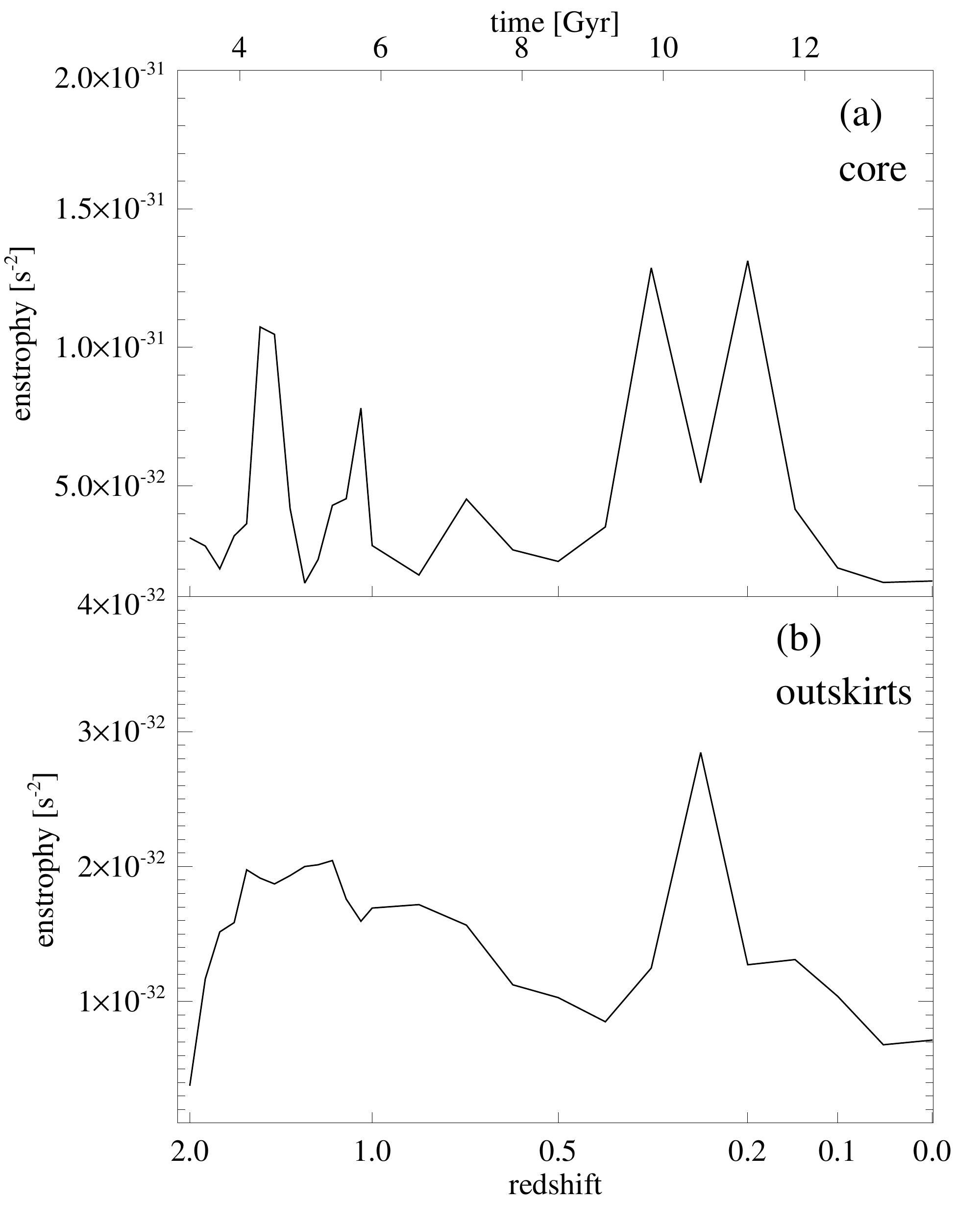}
  \caption{Top panel ($a$): time evolution in the cluster core of the mass-weighted enstrophy in the run $OD4$+.  Bottom panel (b): same as upper one, but for the cluster outskirts. }
  \label{vort-plot}
\end{figure}

In Figure \ref{vort-plot}, the evolution of enstrophy $\mathcal{E}$ in the cluster core and outskirts is shown. This quantity (see also Section \ref{amr}) is defined from the vorticity magnitude as
\begin{equation}
\mathcal{E} = \frac{1}{2} \omega^2 \, \,.
\label{enstrophy}
\end{equation}
At low redshift the enstrophy has peaks at the same times as the internal and SGS turbulence energy (Figure \ref{energy-shells}) both in the core and the outskirts. Analogously to those quantities, it appears therefore sensitive to the stirring on the smallest resolved scales. The correlation between $\mathcal{E}$ and $e_\text{t}$ is expected, as demonstrated by \citet{isn11}. Different than in the time evolution of $e_\text{t}$, the two low-redshift maxima of $\mathcal{E}$ appear of similar magnitude. Moreover, other boosts of $\mathcal{E}$ are visible at higher redshift and are related to the early cluster buildup. The magnitude of the enstrophy peaks appears to correlate only weakly with the masses of the involved merging structures. 

\begin{figure}
  \includegraphics[width=\columnwidth]{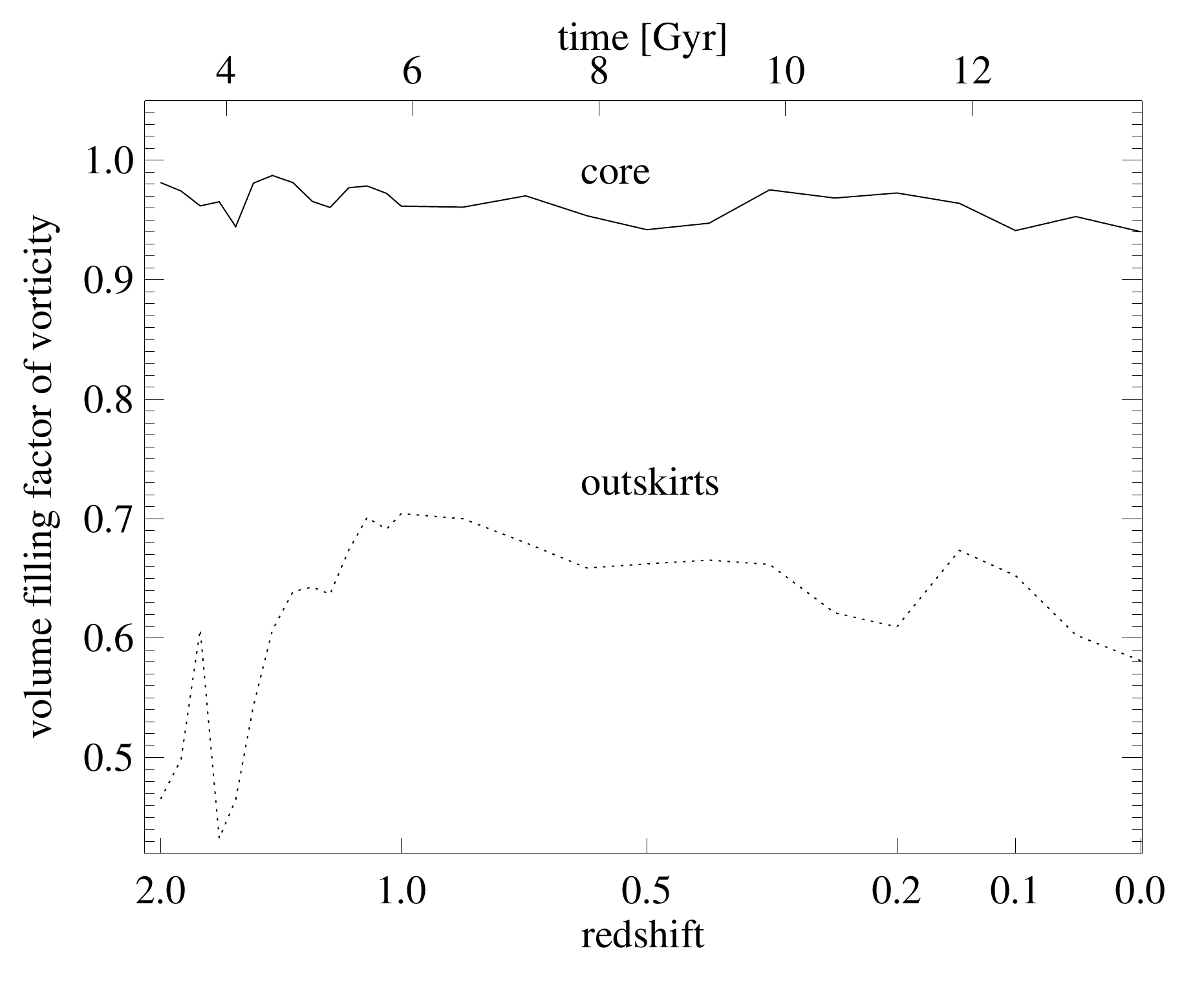}
  \caption{Time evolution of the vorticity volume filling factor (definition in the text) for the cluster core (solid line) and the outskirts (dotted line) for the run $OD4$+. }
  \label{covering}
\end{figure}

Another use of the vorticity is in the characterisation of the turbulent flow. Complementary to its intensity, an important property of turbulence in and around galaxy clusters is its volume filling factor $f_\omega$. It is difficult to find an objective criterion for the definition of this quantity and here, mainly for sake of consistency with previous works, the definition of \citet{krc07} is adopted, namely a cell $i$ is flagged as "turbulent" if the vorticity in the cell
\begin{equation}
\omega_i > N / t_\text{age}(z)  \, \,,
\label{thvort} 
\end{equation}
where $t_\text{age}(z)$ is the age of the universe at redshift $z$ and $N$ is a free parameter. According to this equation, at redshift $z$ we define as turbulent the gas that, within the cluster lifetime (represented by $t_\text{age}(z)$), has a sufficiently large number $N$ of eddy turnovers. At $z=0$ this definition is equivalent to $\omega_i > N H_0$ \citep{m14}. For a better comparison with the cited analyses, we set $N = 10$. The volume filling factor $f_\omega$ in a given region is the volume fraction where the condition expressed by Equation \ref{thvort} is fulfilled.
 
In Figure \ref{covering}, the time evolution of this quantity is shown for the cluster core and the outskirts (defined as in Section \ref{shells}). The filling factor is substantial in the cluster core, where according to the definition above it is always above  $90\%$. A small increase can be seen during the merger events from $z=0.515$. The volume filling factor is quite remarkable also in the outskirts, where it is mostly larger than $60\%$. In the outskirts the evolution at low redshift is decreasing, closely resembling the evolution of $e_\text{t}$ (Figure \ref{energy-shells}). This is likely to be related to the merger history of the cluster, with the last substantial mergers around $z=0.275$ and the subsequent decay of small-scale turbulence.

\begin{table}
\caption{Velocities computed from the mass-weighted energy data in Figure \ref{energy-shells}, for the run $OD4$+, distinguishing between "turbulent'' gas (second and third column) or ''non-turbulent'' gas (fourth and fifth column). The reported velocities are $v$ (computed from  $e_\text{kin}$) and $q$ (from  $e_\text{t}$). The velocities are reported in the outskirts at $z=0$ and $0.275$.}
\centering
\begin{tabular}{lcccc}
\hline
 & $v$ (turb) & $q$ (turb) & $v$ (no turb) & $q$ (no turb) \\ 
 & $[$km\,s$^{-1}]$ & $[$km\,s$^{-1}]$ & $[$km\,s$^{-1}]$ & $[$km\,s$^{-1}]$ \\ 
\hline
outskirts & & & \\
$z=0$ & 1030  & 40.0  & 845 & 19.7 \\
$z=0.275 $  & 1307  & 88.0  & 1142  & 39.4 \\
\label{vort-diff}
\end{tabular}
\end{table}

In Table \ref{vort-diff} one can see the differences of the gas velocities on different scales, when computed in cells within the "turbulent'' volume or outside of it, according to the definition in Equation \ref{thvort}. Since most of the core volume is turbulent, we limit the comparison to the more significant case of the outskirts. The values of $v$ and $q$ in the turbulent gas are very similar to the ones in Table \ref{turbvel}. In the non-turbulent gas the values of $v$ are between $10\%$ and $20\%$ smaller than in the turbulent gas, whereas for $q$ the difference is more pronounced. It is therefore evident that $q$ is a more sensitive diagnostic of the turbulent state of the flow than $v$. This can be understood by considering that $q$ correlates well with the vorticity of the flow \citep{isn11}, and that $v$ includes a laminar velocity component (bulk flow) besides the turbulent one.

\subsection{Diagnostics of resolved and unresolved turbulence: radial profiles}
\label{turb}

In Section \ref{shells} several variable definitions, useful for characterising the properties of turbulence in the ICM and in the cluster outer regions have been introduced. In this way, it has been verified that $e_\text{t}$ is a good diagnostic of turbulence on small length scales. In this Section we want to compare it with another typical indicator, often employed in simulations to characterise the gas flow, namely the mass-weighted  root mean square (henceforth rms) gas velocity $v_\text{rms}$. This variable is computed in radial profiles around the cluster centre and defined as 

\begin{figure}
 \includegraphics[width=\columnwidth]{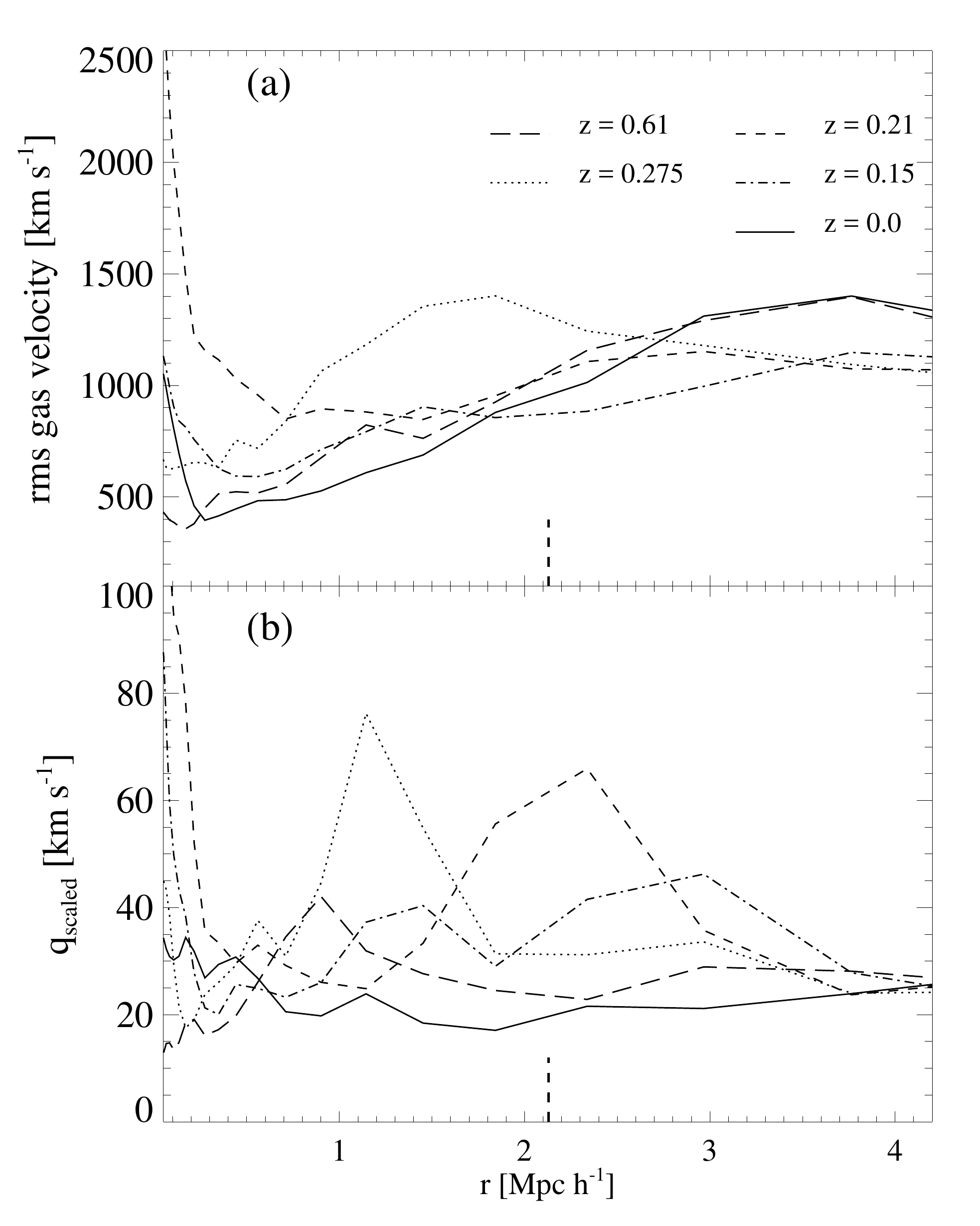}
  \caption{Top panel $(a)$: radial profiles of the rms gas velocity $v_\text{rms}$ (defined in Equation \ref{vrms}) for the simulation $OD4$+, at the times indicated by the respective redshifts in the labels. Bottom panel $(b)$: same as above, but the profiles of $q_\text{scaled}$ are shown. In both panels, the length on the $x$-axis is set in such a way to enclose $2\,R_\text{vir}$ at $z=0$. The vertical, dashed line marks the virial radius at $z=0$, and is a good approximation of it from $z=0.35$.}
  \label{ev-sigma}
\end{figure}

\begin{equation}
v_\text{rms}(r) = \sqrt{\frac{\sum_i m_i \sum_j(v_{i,j} - \langle v_j(r) \rangle)^2}{\sum_i m_i}}\,\,,
\label{vrms}
\end{equation}
where $m_i$ is the mass contained in the cell $i$, with the summation on $i$
computed over the cells belonging to the spherical shell centred on $r$, and that on $j$ over the spatial directions; $\langle \bmath{v}(r) \rangle$ is the average velocity in the shell centred at radius $r$. A time sequence of profiles of $v_\text{rms}$ for the run $OD4$+  is presented in Figure \ref{ev-sigma}$a$. For better readability, we included only two profiles in pre-merger and post-merger phase ($z=0.61$ and $z=0$, respectively) and a sequence of profiles following the major merger. Especially in the latter ones one can see boosts of $v_\text{rms}$, within the virial radius, which then slowly decay. The magnitude of $v_\text{rms}$  is of the same order of $v$, 
indicative of the flow on large length scales.

The profiles of $v_\text{rms}$ (and particularly the ones corresponding to relatively quiet late phases of the evolution) grow with increasing distance from the cluster centre by about  a factor of two. This behaviour has been often misinterpreted as a sign of more intense turbulent flow in the cluster outskirts. Indeed, as also observed by \citet{valda11}, the meaning of the radial profiles of  $v_\text{rms}$ is ambiguous, because no filtering scale for turbulence is explicitly included in Equation (\ref{vrms}), and this is basically equivalent to the size of the spherical shell used for the computation. The outermost shells are therefore dominated by the large-scale (order of Mpc) bulk flows and do not convey information on the turbulent gas motions on smaller scales.  

It is interesting to distinguish which fraction of $v_\text{rms}$ is radial, namely caused by infall of gas driven by gravitational accretion (e.g.~\citealt{kh10,fss11}), and which remaining  component is most genuinely contributing  to turbulence motions. For example, \citet{nln14b} show that the radial component of the rms gas velocity grows in the outskirts (see also \citealt{in08}, for a radial velocity profile in a relaxed cluster). This could explain the growth of $v_\text{rms}$ in the cluster outer region, as due mainly to accretion. This indication should be read with a cautionary remark: during a major merger, like the crucial phases described in this work, a strict  spherical symmetry is hard to hold, and the idea of isolating a radial component in $v_\text{rms}$ is questionable. 

By its very nature, the turbulent SGS energy is a scale-dependent quantity. In some problems, it would be quite helpful to define a quantity where the dependency on the spatial resolution (more technically, on the AMR level of the grid) is properly accounted and tentatively corrected. To achieve this, \citet{mis09} define a "scaled" SGS turbulence energy, obtained by assuming a local Kolmogorov scaling for the turbulence in the ICM and modifying the value of the SGS turbulence energy in a grid cell depending on its AMR level. According to the assumption of Kolmogorov scaling in an incompressible turbulent flow, 
\begin{equation}
v_1 = v_2 \left(\frac{l_{\Delta,1}}{l_{\Delta,2}}\right)^{1/3} \, \,,
\label{kolmo_v}
\end{equation}
where $v_1$ and $v_2$ are turbulence velocities on different length scales $l_{\Delta, 1}$ and $l_{\Delta, 2}$, both within the inertial range of turbulence\footnote{In contrast to the Kolmogorov (incompressible) turbulence, where the scaling follows $v \sim l^{1/3}$, highly supersonic, Burgers (compressible) turbulence follows a stronger scaling with $v \sim l^{1/2}$ \citep{f13}.}.
In our numerical scheme, $v_1$ and $v_2$ can now be interpreted as turbulent velocities on subgrid scales, and $l_{\Delta,1}$ and $l_{\Delta,2}$ as grid resolutions corresponding to the levels of refinement $l_1$ and $l_2$. Using this interpretation of Equation (\ref{kolmo_v}), we implicitly assume that this statistical relation holds locally. 

Based on Equation (\ref{kolmo_v}) and on the previous assumption, we see that the SGS turbulent
energies on two different levels of refinement $l_1$ and $l_2$ with cell size $l_{\Delta,1}$
and $l_{\Delta,2}$, respectively, are related by:
\begin{equation}
\frac{e_\text{t,1}}{e_\text{t,2}}=\frac{v_1^2}{v_2^2}\sim
\left(\frac{l_{\Delta,1}}{l_{\Delta,2}}\right)^{2/3} \, \,.
\label{qscale}
\end{equation}
Thus, taking  $l_{\Delta,\text{max}}$ as the effective spatial resolution at the finest AMR level $l_\text{max}$, we define  
\begin{equation}
\begin{split}
e_\text{t,scaled} \equiv e_{\text{t}, l_\text{max}}= e_\text{t} \left(\frac{l_{\Delta, \text{max}}}{l_{\Delta}}\right)^{2/3}= \\
= e_\text{t} \left(\frac{2^{l}}{2^{l_\text{max}}}\right)^{2/3} = e_\text{t} \times 2^{\frac{2}{3}(l-l_\text{max})}
\end{split}
\label{escale}
\end{equation}
as the turbulence SGS energy that can be computed from $e_\text{t}$ based on Equation (\ref{qscale}), if one uses a Kolmogorov scaling from the generic length scale $l_{\Delta}$ to $l_{\Delta, \text{max}} < l_{\Delta}$. Here one makes use of the relation between the coarse grid resolution $l_{\Delta, \text{coarse}}$ (at $l=0$) and the resolution $l_{\Delta}$ at the generic AMR level $l$: $l_{\Delta} = l_{\Delta, \text{coarse}} / 2^l$. In this relation it is assumed that, by setup definition, the resolution changes by a factor of 2 within neighbouring levels of refinement.

To provide a comparison term, in Figure \ref{ev-sigma}$b$ we present the profiles of the velocity $q_\text{scaled}$, computed from the scaled SGS turbulence energy $e_\text{t,scaled}$ as $q_\text{scaled} = \sqrt{2 e_\text{t,scaled}}$. In this definition, only turbulence on small (in this case, sub-grid) scales contributes, and the bulk flow on larger length scales is filtered out.
One  can clearly see that the profiles of $q_\text{scaled}$ are relatively flat, up to large distances from the cluster centre.
Moreover, this variable is quite useful to highlight merger-induced turbulence, like the stirring caused by the merger shock, visible as the peak at $r=1.2$\,Mpc\,$h^{-1}$ at $z=0.275$, moving outwards in the two subsequent times. The displacement and the decrease of the peak in $q_\text{scaled}$ in different profiles is related to the decrease in time of $e_\text{t}$ seen in the outskirts, in Figure \ref{energy-shells}. One can therefore conclude from this figure that, fixing a given length scale (in our case, of the order of the effective spatial resolution of our simulation sample) the SGS turbulence energy does not systematically increase in the cluster outskirts, as also shown by \citet{valda11}, \citet{vrb12} and \citet{sab14} for other diagnostics of turbulence on small length scales.

\section{Discussion}
\label{discussion}

\subsection{Turbulence in galaxy clusters}
\label{main-result}

A merger event injects turbulence in the ICM. One of the main results of this work is to show that 
this stirring mechanism dominates not only in the cluster core, but also in the outer regions up to several virial radii. Although at first it appears intuitive, this is in contrast with earlier, less resolved simulation results by \citet{isn11}, and to the theoretical analysis of the outskirts at low redshift by \citet{clf11}, which considers the effect of gas accretion onto the clusters, but not their merger activity. In both those studies, the energy in turbulent motions in the cluster peripheries increases with time to the current epoch. The same increase of turbulence in the outskirts, tracked by vorticity, is also observed in the recent simulation of \citet{vjb16}. It is hard to a attempt a comparison on this point with that work, which refers to a single cluster undergoing a major merger but with smaller mass (and thus different energy budget involved) than this study. The analysis of a larger simulation sample would be beneficial on this problem.
 
We have seen in Section \ref{shells} that turbulence diagnostics on different length scales do not have the same time evolution. In Figure \ref{energy-shells} the values of $e_\text{kin}$ 
start increasing earlier than $e_\text{t}$ as a consequence of the major merger. We interpreted it as an effect of stirring on large scales, which only in a later phase is imprinted on small (subgrid) scales. Interestingly, in the core $e_\text{kin}$ 
starts increasing even \textit{before} the major merger (according to the {\sc hop} tool, the merging structure is identified as a single cluster at $z=0.427$). Moreover, a series of minor merger events follow the major one (Section \ref{features}). The resulting picture is in agreement with our knowledge of the cosmic large-scale structure (cf.\ \citealt{vbg11,m15}): a cluster accreting along a filament is accompanied by minor clumps and by the filament gas itself. The dissipation of turbulence in clusters is therefore a more complex process than the decay in forced, idealised simulations: besides the main driver of turbulence (in our case, the major merger), there are further minor stirring agents anticipating and following it. 

A corollary of the complex process described above is that it is difficult to provide a unique and consistent definition of the turbulence dissipation time in clusters. The duration of the turbulence decay (i.e.\ how long the cluster appears turbulent) depends on multiple driving events. Especially from Figure \ref{vort-plot} one can visually estimate that the dissipation time of a single driving event is less than $2$\,Gyr. However, during a merger the turbulent phase can be much longer.
As an example in use of the tools presented in this work, we estimate here the turbulence decay time $t_\text{decay}$ in the core after the double merger as the time it takes for $e_\text{t}$ to decrease back to the value it had at  $z=0.427$ (when the major merger takes place). This occurs at $z=0.15$, resulting in $t_\text{decay} = 2.7$\,Gyr in the cluster core.
In the outskirts the estimate is complicated by the fact that the merger does not occur in that region. A visual inspection of Figure \ref{energy-shells} suggests that an appropriate choice for $t_\text{decay}$ is the time between $z =0.35$ (corresponding to the peak of $e_\text{kin}$) and $z=0$, namely $4.0$\,Gyr.  Just to provide the comparison with a typical timescale related to turbulence, the turbulence eddy turnover time is $t_\text{eddy} = L / v$, where $L$ is the integral length scale (in the cluster case, of the order of $500$\,kpc\,$h^{-1}$) and $v$ is the typical velocity on the scale $L$, around $1000$\,km\,s$^{-1}$ from Table \ref{turbvel}. The eddy turnover time in the ICM is therefore of the order of $0.7$\,Gyr, and can be better compared with the dissipation time of a single driving event, estimated above.

\citet{pim11} noticed that the relatively long decay time for turbulence in clusters can be problematic to reconcile with the statistics of observed radio halos, in particular with the sharp bimodality between clusters hosting halos or not \citep{bvd07,bcd09}. 
As a possible solution, the acceleration efficiency in the turbulent re-acceleration model \citep{bl11} should be a steep function of the turbulent kinetic energy. Recently, \citet{cbc16} performed an explorative study on statistics of radio halos and fraction of merging galaxy clusters, and put constraints on the timescale of merger-induced disturbance which are similar to the turbulence decay timescale derived in this work.

Although the average values of turbulent velocity are similar in the core and the outskirts at all length scales (as seen in Table \ref{turbvel} and discussed in Section \ref{turb}), the kinetic energy content in the latter regions is larger compared to the internal energy, because these regions have colder gas than the core. This is also clearly shown in terms of the Mach number of the turbulent flow on large length scales, defined here for simplicity as 
\begin{equation}
\mathcal{M}_\text{turb} = \frac{v}{c_\text{s}} = \frac{v}{\sqrt{\gamma (\gamma -1) e_\text{int}}} \, \,,
\label{machturb}
\end{equation}
where $v$ is computed from $e_\text{kin}$ and reported in Table \ref{turbvel}, $c_\text{s}$ is the sound speed, $\gamma = 5/3$ and the value of $e_\text{int}$ is the mass-weighted average computed as in Figure \ref{energy-shells}. The Mach number of the flow is smaller in the cluster core than in the outskirts. The flow is mildly subsonic in the core ($\mathcal{M}_\text{turb} = 0.52$ at $z=0$, $0.75$ at $z=0.350$), and more compressible in the outskirts ($\mathcal{M}_\text{turb} = 1.30$ at $z=0$, $1.32$ at $z=0.275$). Both the velocity values and the Mach numbers are very similar to the recent analysis on thermal Sunyaev-Zel'dovich (SZ) fluctuations in the Coma cluster, performed by \citet{kg16}. Recent results \citep{zdb15} have drawn attention to a point often only hinted (e.g.~\citealt{pim11,ib12}), namely the role of gas and substructures, accreting along cosmic filaments, in driving turbulence (see also \citealt{kh10}, for a more general discussion of accretion as ubiquitous driver of turbulence on a wide range of scales). In a detailed study of this process, a refinement strategy like the one used in this work is mandatory, if one wants to correctly capture the evolution of mildly overdense objects and their role in the cluster energy budget. The interactions between the filaments and the outgoing merger shocks can shape the latter in a way that can reproduce complex observed morphologies, like for example the Toothbrush relic \citep{wri12}. Without any attempt to compare with those data and their interpretation, the straight appearance of the upper merger shock at $z=0.275$ and $0.21$ (Figure \ref{temp-evolution}) is suggestive.

\subsection{Flow in the outskirts and connection to radio relics}
\label{relics}

As mentioned in the Introduction, there is recent evidence that the acceleration model in relics is in tension with the simple DSA, and that one should invoke a pre-existing CR population, either produced by earlier stirring events or ejected by active galaxies \citep{smb15,ssr16}. Other works like \citet{fty15} study a somewhat complementary scenario, namely turbulence driven in the region downstream the propagating merger shock (cf.~\citealt{pim11}), and subsequent re-acceleration of CRs. We notice however that simple theoretical arguments \citep{ib12} show that the flow in the post-shock region cannot sustain fully developed turbulence.   

In principle, the propagation of shocks can expose the upstream conditions of the ICM, and its inhomogeneities. The volume filling factor of the turbulent flow and its turbulent energy could then be linked, through appropriate modelling, with the properties and distribution of a possible seed CR population, that can be re-accelerated during merger events. Turbulence might play a similar role also for the amplification of an upstream magnetic field, further noticing that the driving mechanism (consisting of a mixture of solenoidal or compressive modes) can greatly affect the dynamo action \citep{fcs11,ssk13}. Concerning the spatial distribution of the turbulent flow, in Figure \ref{covering} it was seen that its volume filling factor $f_\omega$ (here defined by a threshold in vorticity equivalent to about $10$ turnovers during the cluster lifetime, see Equation \ref{thvort}) is larger than $60\%$ in the outskirts at $z=0$, and approaches unity in the core. These findings compare well with the high volume-filling fraction of turbulent flows resolved in the simulations by \citet{m14} and \citet{vbg10} and, on the other hand, show that a merger-driven shock has a non-negligible probability of interaction with a medium having a low turbulence energy content.

This whole argument suggests that the fraction of turbulent flow in the outskirts and the level of turbulence would be interesting additions in the models of radio relics for future simulations. These points have never been explicitly considered in previous studies \citep{hby08,bbw12,sxh13}.

\subsection{Caveats and limitations}
\label{caveats}

After having shown the usefulness of our simulation approach, one should however observe that not in all problems a high resolution level in the cluster outer regions is strictly required. As shown by \citet{mis09}, many general properties like radial profiles of thermodynamical variables do not critically depend on the resolution strategies tested in this work. The same is true for variables averaged on relatively large volumes, like the kinetic energy definition introduced in Section \ref{shells}. On the other hand, if one is interested in turbulence injected in environments with low overdensity like post-shock regions in the outskirts or cosmic filaments, resolving small-scale flow requires an AMR approach similar to ours, or resorting to other strategies like sufficiently fine static grids \citep{m14}. 

Moreover, in this work we did not take into account additional physics like radiative cooling and AGN feedback. While the latter is certainly a stirring agent on length scales of the order of $10$\,kpc \citep{vrb12,mbs16}, cooling tends to form denser substructures, which are more effective in producing turbulent wakes in their motion (e.g.~\citealt{ivb13}). Both effects are worth being considered in future work.

Another limitation of this work is its focus on one single galaxy cluster realisation. While a large cluster sample would be certainly beneficial, we tried to critically guide our discussion to general properties of major mergers, rather than to the peculiarities of the simulated case. Some variance in the presented results can be retrieved by the comparison with other single clusters presented in the literature \citep{in08,mis09,m14,m15,vjb16}. 

A recent work by \citet{zcs14} revives an idea which was proposed earlier (e.g.~\citealt{dc05}), namely the role of turbulent heating to offset radiative cooling in galaxy clusters hosting a cooling flow. Our simulations neither model radiative cooling nor AGN outflows, but it should be stressed again that the SGS turbulence energy is a suitable variable to evaluate the role of turbulence on small length scales, and that future studies on this problem could profit from it.

\section{Summary and conclusions}
\label{conclusions}

In this work we present the analysis of a suite of cosmological grid-based simulations, following the evolution of a galaxy cluster undergoing a major merger. In the simulations, a subgrid scale model for the computation of unresolved turbulent energy is used. The runs differ from each other in their AMR strategy; in particular, the properties of a run using refinement based on regional variability of  vorticity are compared with the widely used standard method of refinement based on overdensity. Here we focus on studying the stirring of turbulence in the cluster outskirts and provide a refinement criterion that adequately captures the turbulence in the outskirts. Our main conclusions are summarised as follows:
\begin{enumerate}

\item The refinement based on vorticity is suitable for reaching a good level of resolution outside the cluster core, when compared with strategies based only on gas and DM overdensity. The computational volume refined at the maximum AMR level in run $OD4$+ is a factor of $5.3$ larger than in run $OD2$, performed using the most permissive refinement threshold based on DM and gas overdensity (see Table \ref{tab1sims} for the naming convention of the simulations). Similarly, the volume-weighted average spatial resolution at $r=2\,R_\text{vir}$ in run $OD4$+ is a factor of $1.74$ better than in run $OD2$.
The simulation employing refinement on vorticity is able to better resolve underdense structures at the cluster periphery, like cosmic filaments, and the small-scale stirring associated with the gas inflow onto the ICM. This performance of mesh refinement cannot be obtained with any feasible combination of thresholds of the overdensity criteria only (cf.~Figure \ref{amr-profile}).

\item The turbulence SGS model provides a quantity, the SGS turbulence specific energy $e_\text{t}$, which is a useful indicator of small-scale turbulence and its time evolution during a merger event, in combination with the kinetic energy used as a large-scale diagnostic. This variable is a viable alternative to other approaches of characterising turbulence in galaxy clusters, like the ones in Sections \ref{shells}, \ref{vorticity} and \ref{turb}, and to other filtering methods \citep{lkn09,mis09,vbg11,vrb12,vjb16}. Cluster radial profiles of this quantity show a good correlation with boosts associated with the propagation of a merger shock, and at large distances from the cluster centre are not affected by large-scale bulk flows like the rms gas velocity $v_\text{rms}$.

\item The evolution of the energy budget of the cluster studied
is dominated by the major merger event, not only in the cluster core, but well beyond the virial radius (our analysis has been extended to $2\,R_\text{vir}$, or $4.3$\,Mpc\,$h^{-1}$ at $z=0$), where the merger shock propagates, heating the gas and injecting turbulence in the flow. The timescale for the decay of turbulence is of the order of several $10^9$\,years, because of the complex features of the structure buildup,  involving multiple submergers at different scales. The evolution of the turbulence energy in the outskirts does not show any increase at low redshift caused by accretion of pristine gas, as previously suggested \citep{isn11,clf11}. A firmer conclusion on this point will require the study of more simulations with clusters of different dynamical evolution.

\item On length scales of $10$\,kpc\,$h^{-1}$, of the order of the resolution scale of the simulation, and of the scale best probed by our turbulence SGS model, the turbulence velocity (represented by $q_\text{scaled}$, Section \ref{turb}) is similar both in the cluster core and in the outskirts (Figure \ref{ev-sigma}$b$), but the flow is more compressible in the outskirts, because of the radially decreasing thermal energy profile. 

\item The volume filling factor of flow with vorticity larger than $10/t_\text{age}(z)$ is around $60\%$ at low redshift in the cluster outskirts. It is speculated that the volume fraction and the energy content of turbulent flow can be relevant for the theory of radio relics: useful estimates of magnetic field and pre-accelerated CR populations can be derived, through adequate modelling, from the preshock conditions.

\end{enumerate}

\section*{acknowledgements}
L.I.~thanks W.~Schmidt and J.~Niemeyer for their primary contribution in the development of the turbulence SGS model. The numerical simulations were carried out on {\it SuperMUC} of the Leibniz Supercomputing Centre in Garching (Germany), in the framework of the project {\it pr95he}. C.F.~acknowledges funding provided by the Australian Research Council's Discovery Projects (grants {\it DP150104329} and {\it DP170100603}). C.F.~further acknowledges the J\"ulich Supercomputing Centre (grant {\it hhd20}), the Leibniz Rechenzentrum and the Gauss Centre for Supercomputing (grants {\it pr32lo}, {\it pr48pi} and GCS Large-scale project {\it 10391}), the Australian National Computational Infrastructure (grant~{\it ek9}), and the Pawsey Supercomputing Centre with funding from the Australian Government and the Government of Western Australia. R.S.K.~thanks for support from the European Research Council under the European Community's Seventh Framework Programme (FP7/2007-2013) via the ERC Advanced Grant "STARLIGHT: Formation of the First Stars" under the project number 339177. R.S.K.~furthermore acknowledges funding from the Deutsche Forschungsgemeinschaft via SFB 881, "The Milky Way System" (sub-projects B1, B2 and B8) and from the SPP 1573 "Physics of the Interstellar Medium". The {\sc enzo} code (http://enzo-project.org) is the product of a collaborative effort of scientists at many universities and US national laboratories. Most of the data analysis and visualisation was performed using the {\tt yt} toolkit \citep{tso11}. 

\bibliography{cluster-index}

\begin{thebibliography}{}
\makeatletter
\relax
\def\mn@urlcharsother{\let\do\@makeother \do\$\do\&\do\#\do\^\do\_\do\%\do\~}
\def\mn@doi{\begingroup\mn@urlcharsother \@ifnextchar [ {\mn@doi@}
  {\mn@doi@[]}}
\def\mn@doi@[#1]#2{\def\@tempa{#1}\ifx\@tempa\@empty \href
  {http://dx.doi.org/#2} {doi:#2}\else \href {http://dx.doi.org/#2} {#1}\fi
  \endgroup}
\def\mn@eprint#1#2{\mn@eprint@#1:#2::\@nil}
\def\mn@eprint@arXiv#1{\href {http://arxiv.org/abs/#1} {{\tt arXiv:#1}}}
\def\mn@eprint@dblp#1{\href {http://dblp.uni-trier.de/rec/bibtex/#1.xml}
  {dblp:#1}}
\def\mn@eprint@#1:#2:#3:#4\@nil{\def\@tempa {#1}\def\@tempb {#2}\def\@tempc
  {#3}\ifx \@tempc \@empty \let \@tempc \@tempb \let \@tempb \@tempa \fi \ifx
  \@tempb \@empty \def\@tempb {arXiv}\fi \@ifundefined
  {mn@eprint@\@tempb}{\@tempb:\@tempc}{\expandafter \expandafter \csname
  mn@eprint@\@tempb\endcsname \expandafter{\@tempc}}}

\bibitem[\protect\citeauthoryear{{Akamatsu} \& {Kawahara}}{{Akamatsu} \&
  {Kawahara}}{2013}]{ak13}
{Akamatsu} H.,  {Kawahara} H.,  2013, \mn@doi [\pasj] {10.1093/pasj/65.1.16},
  \href {http://adsabs.harvard.edu/abs/2013PASJ...65...16A} {65, 16}

\bibitem[\protect\citeauthoryear{{Akamatsu}, {Takizawa}, {Nakazawa},
  {Fukazawa}, {Ishisaki}  \& {Ohashi}}{{Akamatsu} et~al.}{2012}]{atn12}
{Akamatsu} H.,  {Takizawa} M.,  {Nakazawa} K.,  {Fukazawa} Y.,  {Ishisaki} Y.,
   {Ohashi} T.,  2012, \pasj, 64, 67

\bibitem[\protect\citeauthoryear{{Avestruz}, {Nagai}, {Lau}  \&
  {Nelson}}{{Avestruz} et~al.}{2015}]{anl14}
{Avestruz} C.,  {Nagai} D.,  {Lau} E.~T.,   {Nelson} K.,  2015, \mn@doi [\apj]
  {10.1088/0004-637X/808/2/176}, 808, 176

\bibitem[\protect\citeauthoryear{{Basu}, {Sommer}, {Erler}, {Eckert}, {Vazza},
  {Magnelli}, {Bertoldi}  \& {Tozzi}}{{Basu} et~al.}{2016}]{bse16}
{Basu} K.,  {Sommer} M.,  {Erler} J.,  {Eckert} D.,  {Vazza} F.,  {Magnelli}
  B.,  {Bertoldi} F.,   {Tozzi} P.,  2016, \mn@doi [\apjl]
  {10.3847/2041-8205/829/2/L23}, 829, L23

\bibitem[\protect\citeauthoryear{{Beresnyak} \& {Miniati}}{{Beresnyak} \&
  {Miniati}}{2016}]{bm15}
{Beresnyak} A.,  {Miniati} F.,  2016, \mn@doi [\apj]
  {10.3847/0004-637X/817/2/127}, 817, 127

\bibitem[\protect\citeauthoryear{{Berger} \& {Colella}}{{Berger} \&
  {Colella}}{1989}]{bc89}
{Berger} M.~J.,  {Colella} P.,  1989, Journal of Computational Physics, 82, 64

\bibitem[\protect\citeauthoryear{{Bonafede}, {Feretti}, {Murgia}, {Govoni},
  {Giovannini}, {Dallacasa}, {Dolag}  \& {Taylor}}{{Bonafede}
  et~al.}{2010}]{bfm10}
{Bonafede} A.,  {Feretti} L.,  {Murgia} M.,  {Govoni} F.,  {Giovannini} G.,
  {Dallacasa} D.,  {Dolag} K.,   {Taylor} G.~B.,  2010, \mn@doi [\aap]
  {10.1051/0004-6361/200913696}, 513, A30

\bibitem[\protect\citeauthoryear{{Bonafede}, {Dolag}, {Stasyszyn}, {Murante}
  \& {Borgani}}{{Bonafede} et~al.}{2011}]{bds11}
{Bonafede} A.,  {Dolag} K.,  {Stasyszyn} F.,  {Murante} G.,   {Borgani} S.,
  2011, \mn@doi [\mnras] {10.1111/j.1365-2966.2011.19523.x}, 418, 2234

\bibitem[\protect\citeauthoryear{{Bonafede} et~al.,}{{Bonafede}
  et~al.}{2012}]{bbw12}
{Bonafede} A.,  et~al., 2012, \mn@doi [\mnras]
  {10.1111/j.1365-2966.2012.21570.x}, 426, 40

\bibitem[\protect\citeauthoryear{{Botteon}, {Gastaldello}, {Brunetti}  \&
  {Kale}}{{Botteon} et~al.}{2016}]{bgb16}
{Botteon} A.,  {Gastaldello} F.,  {Brunetti} G.,   {Kale} R.,  2016, \mn@doi
  [\mnras] {10.1093/mnras/stw2089}, 463

\bibitem[\protect\citeauthoryear{{Braun}, {Schmidt}, {Niemeyer}  \&
  {Almgren}}{{Braun} et~al.}{2014}]{bsn14}
{Braun} H.,  {Schmidt} W.,  {Niemeyer} J.~C.,   {Almgren} A.~S.,  2014, \mn@doi
  [\mnras] {10.1093/mnras/stu1119}, 442, 3407

\bibitem[\protect\citeauthoryear{{Br{\"u}ggen}, {Bykov}, {Ryu}  \&
  {R{\"o}ttgering}}{{Br{\"u}ggen} et~al.}{2012}]{bbr12}
{Br{\"u}ggen} M.,  {Bykov} A.,  {Ryu} D.,   {R{\"o}ttgering} H.,  2012, \mn@doi
  [\ssr] {10.1007/s11214-011-9785-9}, 166, 187

\bibitem[\protect\citeauthoryear{{Brunetti} \& {Jones}}{{Brunetti} \&
  {Jones}}{2014}]{bj14}
{Brunetti} G.,  {Jones} T.~W.,  2014, \mn@doi [International Journal of Modern
  Physics D] {10.1142/S0218271814300079}, 23, 1430007

\bibitem[\protect\citeauthoryear{{Brunetti} \& {Lazarian}}{{Brunetti} \&
  {Lazarian}}{2011}]{bl11}
{Brunetti} G.,  {Lazarian} A.,  2011, \mn@doi [\mnras]
  {10.1111/j.1365-2966.2010.17457.x}, 410, 127

\bibitem[\protect\citeauthoryear{{Brunetti}, {Venturi}, {Dallacasa}, {Cassano},
  {Dolag}, {Giacintucci}  \& {Setti}}{{Brunetti} et~al.}{2007}]{bvd07}
{Brunetti} G.,  {Venturi} T.,  {Dallacasa} D.,  {Cassano} R.,  {Dolag} K.,
  {Giacintucci} S.,   {Setti} G.,  2007, \mn@doi [\apjl] {10.1086/524037}, 670,
  L5

\bibitem[\protect\citeauthoryear{{Brunetti}, {Cassano}, {Dolag}  \&
  {Setti}}{{Brunetti} et~al.}{2009}]{bcd09}
{Brunetti} G.,  {Cassano} R.,  {Dolag} K.,   {Setti} G.,  2009, \mn@doi [\aap]
  {10.1051/0004-6361/200912751}, 507, 661

\bibitem[\protect\citeauthoryear{{Bryan} \& {Norman}}{{Bryan} \&
  {Norman}}{1998}]{bn98}
{Bryan} G.~L.,  {Norman} M.~L.,  1998, \mn@doi [\apj] {10.1086/305262}, \href
  {http://ads.ari.uni-heidelberg.de/abs/1998ApJ...495...80B} {495, 80}

\bibitem[\protect\citeauthoryear{{Bryan} et~al.,}{{Bryan}
  et~al.}{2014}]{enzo14}
{Bryan} G.~L.,  et~al., 2014, \mn@doi [\apjs] {10.1088/0067-0049/211/2/19},
  211, 19

\bibitem[\protect\citeauthoryear{{Cassano}, {Brunetti}, {Giocoli}  \&
  {Ettori}}{{Cassano} et~al.}{2016}]{cbc16}
{Cassano} R.,  {Brunetti} G.,  {Giocoli} C.,   {Ettori} S.,  2016, \mn@doi
  [\aap] {10.1051/0004-6361/201628414}, 593, A81

\bibitem[\protect\citeauthoryear{{Cavaliere}, {Lapi}  \&
  {Fusco-Femiano}}{{Cavaliere} et~al.}{2011}]{clf11}
{Cavaliere} A.,  {Lapi} A.,   {Fusco-Femiano} R.,  2011, \mn@doi [\aap]
  {10.1051/0004-6361/201015390}, 525, A110

\bibitem[\protect\citeauthoryear{{Clarke}, {Kronberg}  \&
  {B{\"o}hringer}}{{Clarke} et~al.}{2001}]{ckb01}
{Clarke} T.~E.,  {Kronberg} P.~P.,   {B{\"o}hringer} H.,  2001, \mn@doi [\apjl]
  {10.1086/318896}, 547, L111

\bibitem[\protect\citeauthoryear{{Close}, {Pittard}, {Hartquist}  \&
  {Falle}}{{Close} et~al.}{2013}]{cph13}
{Close} J.~L.,  {Pittard} J.~M.,  {Hartquist} T.~W.,   {Falle} S.~A.~E.~G.,
  2013, \mn@doi [\mnras] {10.1093/mnras/stt1788}, 436, 3021

\bibitem[\protect\citeauthoryear{{Colella} \& {Woodward}}{{Colella} \&
  {Woodward}}{1984}]{cw84}
{Colella} P.,  {Woodward} P.~R.,  1984, \mn@doi [Journal of Computational
  Physics] {10.1016/0021-9991(84)90143-8}, 54, 174

\bibitem[\protect\citeauthoryear{{Dennis} \& {Chandran}}{{Dennis} \&
  {Chandran}}{2005}]{dc05}
{Dennis} T.~J.,  {Chandran} B.~D.~G.,  2005, \mn@doi [\apj] {10.1086/427424},
  622, 205

\bibitem[\protect\citeauthoryear{{Dolag} \& {Stasyszyn}}{{Dolag} \&
  {Stasyszyn}}{2009}]{ds09}
{Dolag} K.,  {Stasyszyn} F.,  2009, \mn@doi [\mnras]
  {10.1111/j.1365-2966.2009.15181.x}, \href
  {http://ads.ari.uni-heidelberg.de/abs/2009MNRAS.tmp.1196D} {398, 1678 }

\bibitem[\protect\citeauthoryear{{Dolag}, {Bartelmann}  \& {Lesch}}{{Dolag}
  et~al.}{2002}]{dbl02}
{Dolag} K.,  {Bartelmann} M.,   {Lesch} H.,  2002, \mn@doi [\aap]
  {10.1051/0004-6361:20020241}, 387, 383

\bibitem[\protect\citeauthoryear{{Dubois} \& {Teyssier}}{{Dubois} \&
  {Teyssier}}{2008}]{dt08}
{Dubois} Y.,  {Teyssier} R.,  2008, \mn@doi [\aap]
  {10.1051/0004-6361:200809513}, 482, L13

\bibitem[\protect\citeauthoryear{{Eckert} et~al.,}{{Eckert}
  et~al.}{2012}]{eve12}
{Eckert} D.,  et~al., 2012, \mn@doi [\aap] {10.1051/0004-6361/201118281}, 541,
  A57

\bibitem[\protect\citeauthoryear{{Eisenstein} \& {Hu}}{{Eisenstein} \&
  {Hu}}{1999}]{eh99}
{Eisenstein} D.~J.,  {Hu} W.,  1999, \mn@doi [ApJ] {10.1086/306640}, 511, 5

\bibitem[\protect\citeauthoryear{{Eisenstein} \& {Hut}}{{Eisenstein} \&
  {Hut}}{1998}]{eh98}
{Eisenstein} D.~J.,  {Hut} P.,  1998, \mn@doi [\apj] {10.1086/305535}, 498, 137

\bibitem[\protect\citeauthoryear{{Favre}}{{Favre}}{1969}]{FAVRE1969}
{Favre} A.,  1969, in {Muskhelishvili} N.~I.,  {Grigolyuk} E.~I.,   {Mikhailov}
  G.~K.,  eds, Problems of hydrodynamics and continuum mechanics. Society for
  Industrial and Applied Mathematics, p.~231

\bibitem[\protect\citeauthoryear{{Federrath}}{{Federrath}}{2013}]{f13}
{Federrath} C.,  2013, \mn@doi [\mnras] {10.1093/mnras/stt1644}, 436, 1245

\bibitem[\protect\citeauthoryear{{Federrath} \& {Klessen}}{{Federrath} \&
  {Klessen}}{2012}]{fk12}
{Federrath} C.,  {Klessen} R.~S.,  2012, \mn@doi [\apj]
  {10.1088/0004-637X/761/2/156}, 761, 156

\bibitem[\protect\citeauthoryear{{Federrath}, {Roman-Duval}, {Klessen},
  {Schmidt}  \& {Mac Low}}{{Federrath} et~al.}{2010}]{frk10}
{Federrath} C.,  {Roman-Duval} J.,  {Klessen} R.~S.,  {Schmidt} W.,   {Mac Low}
  M.-M.,  2010, \mn@doi [\aap] {10.1051/0004-6361/200912437}, \href
  {http://ads.ari.uni-heidelberg.de/abs/2010A%26A...512A..81F} {512, A81}

\bibitem[\protect\citeauthoryear{{Federrath}, {Chabrier}, {Schober},
  {Banerjee}, {Klessen}  \& {Schleicher}}{{Federrath} et~al.}{2011a}]{fcs11}
{Federrath} C.,  {Chabrier} G.,  {Schober} J.,  {Banerjee} R.,  {Klessen}
  R.~S.,   {Schleicher} D.~R.~G.,  2011a, \mn@doi [Phys. Rev. Lett.]
  {10.1103/PhysRevLett.107.114504}, 107, 114504

\bibitem[\protect\citeauthoryear{{Federrath}, {Sur}, {Schleicher}, {Banerjee}
  \& {Klessen}}{{Federrath} et~al.}{2011b}]{fss11}
{Federrath} C.,  {Sur} S.,  {Schleicher} D.~R.~G.,  {Banerjee} R.,   {Klessen}
  R.~S.,  2011b, \mn@doi [\apj] {10.1088/0004-637X/731/1/62}, 731, 62

\bibitem[\protect\citeauthoryear{{Ferrari}, {Govoni}, {Schindler}, {Bykov}  \&
  {Rephaeli}}{{Ferrari} et~al.}{2008}]{fgs08}
{Ferrari} C.,  {Govoni} F.,  {Schindler} S.,  {Bykov} A.~M.,   {Rephaeli} Y.,
  2008, \mn@doi [\ssr] {10.1007/s11214-008-9311-x}, \href
  {http://ads.ari.uni-heidelberg.de/abs/2008SSRv..134...93F} {134, 93}

\bibitem[\protect\citeauthoryear{{Finoguenov}, {Sarazin}, {Nakazawa}, {Wik}  \&
  {Clarke}}{{Finoguenov} et~al.}{2010}]{fsn10}
{Finoguenov} A.,  {Sarazin} C.~L.,  {Nakazawa} K.,  {Wik} D.~R.,   {Clarke}
  T.~E.,  2010, \mn@doi [\apj] {10.1088/0004-637X/715/2/1143}, 715, 1143

\bibitem[\protect\citeauthoryear{{Frisch}}{{Frisch}}{1995}]{f95}
{Frisch} U.,  1995, {Turbulence. The legacy of A.N. Kolmogorov}.
Cambridge: Cambridge University Press, \url
  {http://esoads.eso.org/cgi-bin/nph-bib_query?bibcode=1995tlnk.book.....F&db_key=AST}

\bibitem[\protect\citeauthoryear{{Fujita}, {Takizawa}, {Yamazaki}, {Akamatsu}
  \& {Ohno}}{{Fujita} et~al.}{2015}]{fty15}
{Fujita} Y.,  {Takizawa} M.,  {Yamazaki} R.,  {Akamatsu} H.,   {Ohno} H.,
  2015, \mn@doi [\apj] {10.1088/0004-637X/815/2/116}, 815, 116

\bibitem[\protect\citeauthoryear{Germano}{Germano}{1992}]{Germano1992}
Germano M.,  1992, Journal of Fluid Mechanics, 238, 325

\bibitem[\protect\citeauthoryear{{Giacintucci} et~al.,}{{Giacintucci}
  et~al.}{2008}]{gvm08}
{Giacintucci} S.,  et~al., 2008, \mn@doi [\aap] {10.1051/0004-6361:200809459},
  \href {http://esoads.eso.org/abs/2008A%26A...486..347G} {486, 347}

\bibitem[\protect\citeauthoryear{{Govoni}, {Murgia}, {Feretti}, {Giovannini},
  {Dolag}  \& {Taylor}}{{Govoni} et~al.}{2006}]{gmf06}
{Govoni} F.,  {Murgia} M.,  {Feretti} L.,  {Giovannini} G.,  {Dolag} K.,
  {Taylor} G.~B.,  2006, \mn@doi [\aap] {10.1051/0004-6361:20065964}, 460, 425

\bibitem[\protect\citeauthoryear{{Hitomi Collaboration} et~al.,}{{Hitomi
  Collaboration} et~al.}{2016}]{hitomi16}
{Hitomi Collaboration} et~al., 2016, \mn@doi [Nature] {10.1038/nature18627},
  535, 117

\bibitem[\protect\citeauthoryear{{Hockney} \& {Eastwood}}{{Hockney} \&
  {Eastwood}}{1988}]{he88}
{Hockney} R.~W.,  {Eastwood} J.~W.,  1988, {Computer simulation using
  particles}.
Bristol: Hilger, \url {http://adsabs.harvard.edu/abs/1988csup.book.....H}

\bibitem[\protect\citeauthoryear{{Hoeft}, {Br{\"u}ggen}, {Yepes},
  {Gottl{\"o}ber}  \& {Schwope}}{{Hoeft} et~al.}{2008}]{hby08}
{Hoeft} M.,  {Br{\"u}ggen} M.,  {Yepes} G.,  {Gottl{\"o}ber} S.,   {Schwope}
  A.,  2008, \mn@doi [\mnras] {10.1111/j.1365-2966.2008.13955.x}, 391, 1511

\bibitem[\protect\citeauthoryear{{Iapichino} \& {Br{\"u}ggen}}{{Iapichino} \&
  {Br{\"u}ggen}}{2012}]{ib12}
{Iapichino} L.,  {Br{\"u}ggen} M.,  2012, \mn@doi [\mnras]
  {10.1111/j.1365-2966.2012.21084.x}, 423, 2781

\bibitem[\protect\citeauthoryear{{Iapichino} \& {Niemeyer}}{{Iapichino} \&
  {Niemeyer}}{2008}]{in08}
{Iapichino} L.,  {Niemeyer} J.~C.,  2008, \mn@doi [\mnras]
  {10.1111/j.1365-2966.2008.13518.x}, 388, 1089

\bibitem[\protect\citeauthoryear{{Iapichino}, {Adamek}, {Schmidt}  \&
  {Niemeyer}}{{Iapichino} et~al.}{2008}]{ias08}
{Iapichino} L.,  {Adamek} J.,  {Schmidt} W.,   {Niemeyer} J.~C.,  2008, \mn@doi
  [\mnras] {10.1111/j.1365-2966.2008.13137.x}, 388, 1079

\bibitem[\protect\citeauthoryear{{Iapichino}, {Schmidt}, {Niemeyer}  \&
  {Merklein}}{{Iapichino} et~al.}{2011}]{isn11}
{Iapichino} L.,  {Schmidt} W.,  {Niemeyer} J.~C.,   {Merklein} J.,  2011,
  \mn@doi [\mnras] {10.1111/j.1365-2966.2011.18550.x}, 414, 2297

\bibitem[\protect\citeauthoryear{{Iapichino}, {Viel}  \& {Borgani}}{{Iapichino}
  et~al.}{2013}]{ivb13}
{Iapichino} L.,  {Viel} M.,   {Borgani} S.,  2013, \mn@doi [\mnras]
  {10.1093/mnras/stt611}, 432, 2529

\bibitem[\protect\citeauthoryear{{Ichinohe}, {Werner}, {Simionescu}, {Allen},
  {Canning}, {Ehlert}, {Mernier}  \& {Takahashi}}{{Ichinohe}
  et~al.}{2015}]{iws14}
{Ichinohe} Y.,  {Werner} N.,  {Simionescu} A.,  {Allen} S.~W.,  {Canning}
  R.~E.~A.,  {Ehlert} S.,  {Mernier} F.,   {Takahashi} T.,  2015, \mn@doi
  [\mnras] {10.1093/mnras/stv217}, 448, 2971

\bibitem[\protect\citeauthoryear{{Kang} \& {Ryu}}{{Kang} \& {Ryu}}{2011}]{kr11}
{Kang} H.,  {Ryu} D.,  2011, \mn@doi [\apj] {10.1088/0004-637X/734/1/18}, 734,
  18

\bibitem[\protect\citeauthoryear{{Kang}, {Ryu}, {Cen}  \& {Ostriker}}{{Kang}
  et~al.}{2007}]{krc07}
{Kang} H.,  {Ryu} D.,  {Cen} R.,   {Ostriker} J.~P.,  2007, \mn@doi [\apj]
  {10.1086/521717}, 669, 729

\bibitem[\protect\citeauthoryear{{Kang}, {Ryu}  \& {Jones}}{{Kang}
  et~al.}{2012}]{krj12}
{Kang} H.,  {Ryu} D.,   {Jones} T.~W.,  2012, \mn@doi [\apj]
  {10.1088/0004-637X/756/1/97}, \href
  {http://adsabs.harvard.edu/abs/2012ApJ...756...97K} {756, 97}

\bibitem[\protect\citeauthoryear{{Khatri} \& {Gaspari}}{{Khatri} \&
  {Gaspari}}{2016}]{kg16}
{Khatri} R.,  {Gaspari} M.,  2016, \mn@doi [\mnras] {10.1093/mnras/stw2027},
  463, 655

\bibitem[\protect\citeauthoryear{{Klessen} \& {Hennebelle}}{{Klessen} \&
  {Hennebelle}}{2010}]{kh10}
{Klessen} R.~S.,  {Hennebelle} P.,  2010, \mn@doi [\aap]
  {10.1051/0004-6361/200913780}, 520, A17

\bibitem[\protect\citeauthoryear{{Kolmogorov}}{{Kolmogorov}}{1941}]{k41}
{Kolmogorov} A.,  1941, Akademiia Nauk SSSR Doklady, 30, 301

\bibitem[\protect\citeauthoryear{{Kritsuk}, {Norman}  \& {Padoan}}{{Kritsuk}
  et~al.}{2006}]{knp06}
{Kritsuk} A.~G.,  {Norman} M.~L.,   {Padoan} P.,  2006, \mn@doi [\apjl]
  {10.1086/500688}, 638, L25

\bibitem[\protect\citeauthoryear{{Lau}, {Kravtsov}  \& {Nagai}}{{Lau}
  et~al.}{2009}]{lkn09}
{Lau} E.~T.,  {Kravtsov} A.~V.,   {Nagai} D.,  2009, \mn@doi [\apj]
  {10.1088/0004-637X/705/2/1129}, 705, 1129

\bibitem[\protect\citeauthoryear{{Macario}, {Markevitch}, {Giacintucci},
  {Brunetti}, {Venturi}  \& {Murray}}{{Macario} et~al.}{2011}]{mmg11}
{Macario} G.,  {Markevitch} M.,  {Giacintucci} S.,  {Brunetti} G.,  {Venturi}
  T.,   {Murray} S.~S.,  2011, \mn@doi [\apj] {10.1088/0004-637X/728/2/82},
  728, 82

\bibitem[\protect\citeauthoryear{{Maier}, {Iapichino}, {Schmidt}  \&
  {Niemeyer}}{{Maier} et~al.}{2009}]{mis09}
{Maier} A.,  {Iapichino} L.,  {Schmidt} W.,   {Niemeyer} J.~C.,  2009, \mn@doi
  [\apj] {10.1088/0004-637X/707/1/40}, 707, 40

\bibitem[\protect\citeauthoryear{{Meinecke} et~al.,}{{Meinecke}
  et~al.}{2014}]{mdm14}
{Meinecke} J.,  et~al., 2014, \mn@doi [Nature Physics] {10.1038/nphys2978}, 10,
  520

\bibitem[\protect\citeauthoryear{Meinecke et~al.,}{Meinecke
  et~al.}{2015}]{mtb15}
Meinecke J.,  et~al., 2015, \mn@doi [Proceedings of the National Academy of
  Sciences] {10.1073/pnas.1502079112}, 112, 8211

\bibitem[\protect\citeauthoryear{{Miniati}}{{Miniati}}{2014}]{m14}
{Miniati} F.,  2014, \mn@doi [\apj] {10.1088/0004-637X/782/1/21}, 782, 21

\bibitem[\protect\citeauthoryear{{Miniati}}{{Miniati}}{2015}]{m15}
{Miniati} F.,  2015, \mn@doi [\apj] {10.1088/0004-637X/800/1/60}, 800, 60

\bibitem[\protect\citeauthoryear{{Miniati} \& {Beresnyak}}{{Miniati} \&
  {Beresnyak}}{2015}]{mb15}
{Miniati} F.,  {Beresnyak} A.,  2015, \mn@doi [Nature] {10.1038/nature14552},
  523, 59

\bibitem[\protect\citeauthoryear{{Morandi}, {Nagai}  \& {Cui}}{{Morandi}
  et~al.}{2013}]{mnc13}
{Morandi} A.,  {Nagai} D.,   {Cui} W.,  2013, \mn@doi [\mnras]
  {10.1093/mnras/stt1636}, 436, 1123

\bibitem[\protect\citeauthoryear{{Mukherjee}, {Bicknell}, {Sutherland}  \&
  {Wagner}}{{Mukherjee} et~al.}{2016}]{mbs16}
{Mukherjee} D.,  {Bicknell} G.~V.,  {Sutherland} R.,   {Wagner} A.,  2016,
  \mn@doi [\mnras] {10.1093/mnras/stw1368}, 461, 967

\bibitem[\protect\citeauthoryear{{Nelson}, {Lau}  \& {Nagai}}{{Nelson}
  et~al.}{2014}]{nln14b}
{Nelson} K.,  {Lau} E.~T.,   {Nagai} D.,  2014, \mn@doi [\apj]
  {10.1088/0004-637X/792/1/25}, 792, 25

\bibitem[\protect\citeauthoryear{{O'Shea}, {Bryan}, {Bordner}, {Norman},
  {Abel}, {Harkness}  \& {Kritsuk}}{{O'Shea} et~al.}{2005a}]{obb05}
{O'Shea} B.~W.,  {Bryan} G.,  {Bordner} J.,  {Norman} M.~L.,  {Abel} T.,
  {Harkness} R.,   {Kritsuk} A.,  2005a, in Adaptive Mesh Refinement -- Theory
  and Applications, ed.~T.~Plewa, T.~Linde, V.G.~Weirs (Berlin; New York:
  Springer). p.~341, \url {http://esoads.eso.org/abs/2004astro.ph..3044O}

\bibitem[\protect\citeauthoryear{{O'Shea}, {Nagamine}, {Springel}, {Hernquist}
  \& {Norman}}{{O'Shea} et~al.}{2005b}]{ons05}
{O'Shea} B.~W.,  {Nagamine} K.,  {Springel} V.,  {Hernquist} L.,   {Norman}
  M.~L.,  2005b, \mn@doi [\apjs] {10.1086/432645}, 160, 1

\bibitem[\protect\citeauthoryear{{Ostriker}}{{Ostriker}}{1993}]{o93}
{Ostriker} J.~P.,  1993, \mn@doi [\araa] {10.1146/annurev.aa.31.090193.003353},
  31, 689

\bibitem[\protect\citeauthoryear{{Parrish}, {McCourt}, {Quataert}  \&
  {Sharma}}{{Parrish} et~al.}{2012}]{pmq12}
{Parrish} I.~J.,  {McCourt} M.,  {Quataert} E.,   {Sharma} P.,  2012, \mn@doi
  [\mnras] {10.1111/j.1745-3933.2011.01171.x}, 419, L29

\bibitem[\protect\citeauthoryear{{Paul}, {Iapichino}, {Miniati}, {Bagchi}  \&
  {Mannheim}}{{Paul} et~al.}{2011}]{pim11}
{Paul} S.,  {Iapichino} L.,  {Miniati} F.,  {Bagchi} J.,   {Mannheim} K.,
  2011, \mn@doi [\apj] {10.1088/0004-637X/726/1/17}, 726, 17

\bibitem[\protect\citeauthoryear{{Pinzke}, {Oh}  \& {Pfrommer}}{{Pinzke}
  et~al.}{2013}]{pop13}
{Pinzke} A.,  {Oh} S.~P.,   {Pfrommer} C.,  2013, \mn@doi [\mnras]
  {10.1093/mnras/stt1308}, 435, 1061

\bibitem[\protect\citeauthoryear{{Planck Collaboration} et~al.,}{{Planck
  Collaboration} et~al.}{2014}]{planck14}
{Planck Collaboration} et~al., 2014, \mn@doi [\aap]
  {10.1051/0004-6361/201321529}, \href
  {http://adsabs.harvard.edu/abs/2014A%26A...571A...1P} {571, A1}

\bibitem[\protect\citeauthoryear{{Rasia}, {Tormen}  \& {Moscardini}}{{Rasia}
  et~al.}{2004}]{rtm04}
{Rasia} E.,  {Tormen} G.,   {Moscardini} L.,  2004, \mn@doi [\mnras]
  {10.1111/j.1365-2966.2004.07775.x}, 351, 237

\bibitem[\protect\citeauthoryear{{Reiprich}, {Basu}, {Ettori}, {Israel},
  {Lovisari}, {Molendi}, {Pointecouteau}  \& {Roncarelli}}{{Reiprich}
  et~al.}{2013}]{rbe13}
{Reiprich} T.~H.,  {Basu} K.,  {Ettori} S.,  {Israel} H.,  {Lovisari} L.,
  {Molendi} S.,  {Pointecouteau} E.,   {Roncarelli} M.,  2013, \mn@doi [\ssr]
  {10.1007/s11214-013-9983-8}, 177, 195

\bibitem[\protect\citeauthoryear{{Renaud}, {Kraljic}  \& {Bournaud}}{{Renaud}
  et~al.}{2012}]{rkb12}
{Renaud} F.,  {Kraljic} K.,   {Bournaud} F.,  2012, \mn@doi [\apjl]
  {10.1088/2041-8205/760/1/L16}, 760, L16

\bibitem[\protect\citeauthoryear{{Renaud}, {Bournaud}, {Kraljic}  \&
  {Duc}}{{Renaud} et~al.}{2014}]{rbk14}
{Renaud} F.,  {Bournaud} F.,  {Kraljic} K.,   {Duc} P.-A.,  2014, \mn@doi
  [\mnras] {10.1093/mnrasl/slu050}, 442, L33

\bibitem[\protect\citeauthoryear{{Roncarelli}, {Ettori}, {Borgani}, {Dolag},
  {Fabjan}  \& {Moscardini}}{{Roncarelli} et~al.}{2013}]{reb13}
{Roncarelli} M.,  {Ettori} S.,  {Borgani} S.,  {Dolag} K.,  {Fabjan} D.,
  {Moscardini} L.,  2013, \mn@doi [\mnras] {10.1093/mnras/stt654}, 432, 3030

\bibitem[\protect\citeauthoryear{Ryu, Kang, Cho  \& Das}{Ryu
  et~al.}{2008}]{rkc08}
Ryu D.,  Kang H.,  Cho J.,   Das S.,  2008, \mn@doi [Science]
  {10.1126/science.1154923}, 320, 909

\bibitem[\protect\citeauthoryear{{Scannapieco} \& {Br{\"u}ggen}}{{Scannapieco}
  \& {Br{\"u}ggen}}{2008}]{sb08}
{Scannapieco} E.,  {Br{\"u}ggen} M.,  2008, \mn@doi [\apj] {10.1086/591228},
  686, 927

\bibitem[\protect\citeauthoryear{{Schleicher} et~al.,}{{Schleicher}
  et~al.}{2013}]{sls13}
{Schleicher} D.~R.~G.,  et~al., 2013, \mn@doi [Astronomische Nachrichten]
  {10.1002/asna.201211898}, 334, 531

\bibitem[\protect\citeauthoryear{{Schmidt}}{{Schmidt}}{2015}]{s15}
{Schmidt} W.,  2015, \mn@doi [Living Reviews in Computational Astrophysics]
  {10.1007/lrca-2015-2}, \href
  {http://adsabs.harvard.edu/abs/2015LRCA....1....2S} {1}

\bibitem[\protect\citeauthoryear{{Schmidt} \& {Federrath}}{{Schmidt} \&
  {Federrath}}{2011}]{sf11}
{Schmidt} W.,  {Federrath} C.,  2011, \mn@doi [\aap]
  {10.1051/0004-6361/201015630}, 528, A106

\bibitem[\protect\citeauthoryear{{Schmidt}, {Niemeyer}  \&
  {Hillebrandt}}{{Schmidt} et~al.}{2006a}]{snh06}
{Schmidt} W.,  {Niemeyer} J.~C.,   {Hillebrandt} W.,  2006a, \mn@doi [\aap]
  {10.1051/0004-6361:20053617}, 450, 265

\bibitem[\protect\citeauthoryear{{Schmidt}, {Niemeyer}, {Hillebrandt}  \&
  {R{\"o}pke}}{{Schmidt} et~al.}{2006b}]{snhr06}
{Schmidt} W.,  {Niemeyer} J.~C.,  {Hillebrandt} W.,   {R{\"o}pke} F.~K.,
  2006b, \mn@doi [\aap] {10.1051/0004-6361:20053618}, 450, 283

\bibitem[\protect\citeauthoryear{{Schmidt}, {Federrath}, {Hupp}, {Kern}  \&
  {Niemeyer}}{{Schmidt} et~al.}{2009}]{sfh09}
{Schmidt} W.,  {Federrath} C.,  {Hupp} M.,  {Kern} S.,   {Niemeyer} J.~C.,
  2009, \mn@doi [\aap] {10.1051/0004-6361:200809967}, 494, 127

\bibitem[\protect\citeauthoryear{{Schmidt} et~al.,}{{Schmidt}
  et~al.}{2014}]{sab14}
{Schmidt} W.,  et~al., 2014, \mn@doi [\mnras] {10.1093/mnras/stu501}, 440, 3051

\bibitem[\protect\citeauthoryear{{Schmidt}, {Schulz}, {Iapichino}, {Vazza}  \&
  {Almgren}}{{Schmidt} et~al.}{2015}]{ssi15}
{Schmidt} W.,  {Schulz} J.,  {Iapichino} L.,  {Vazza} F.,   {Almgren} A.~S.,
  2015, Astronomy and Computing, \href
  {http://ads.ari.uni-heidelberg.de/abs/2014arXiv1411.7275S} {9, 49}

\bibitem[\protect\citeauthoryear{{Schober}, {Schleicher}  \&
  {Klessen}}{{Schober} et~al.}{2013}]{ssk13}
{Schober} J.,  {Schleicher} D.~R.~G.,   {Klessen} R.~S.,  2013, \mn@doi [\aap]
  {10.1051/0004-6361/201322185}, 560, A87

\bibitem[\protect\citeauthoryear{{Shen}, {Wadsley}  \& {Stinson}}{{Shen}
  et~al.}{2010}]{sws10}
{Shen} S.,  {Wadsley} J.,   {Stinson} G.,  2010, \mn@doi [\mnras]
  {10.1111/j.1365-2966.2010.17047.x}, 407, 1581

\bibitem[\protect\citeauthoryear{{Shi}, {Komatsu}, {Nelson}  \& {Nagai}}{{Shi}
  et~al.}{2015}]{skn15}
{Shi} X.,  {Komatsu} E.,  {Nelson} K.,   {Nagai} D.,  2015, \mn@doi [\mnras]
  {10.1093/mnras/stv036}, 448, 1020

\bibitem[\protect\citeauthoryear{{Shimwell}, {Markevitch}, {Brown}, {Feretti},
  {Gaensler}, {Johnston-Hollitt}, {Lage}  \& {Srinivasan}}{{Shimwell}
  et~al.}{2015}]{smb15}
{Shimwell} T.~W.,  {Markevitch} M.,  {Brown} S.,  {Feretti} L.,  {Gaensler}
  B.~M.,  {Johnston-Hollitt} M.,  {Lage} C.,   {Srinivasan} R.,  2015, \mn@doi
  [\mnras] {10.1093/mnras/stv334}, 449, 1486

\bibitem[\protect\citeauthoryear{{Simionescu} et~al.,}{{Simionescu}
  et~al.}{2011}]{sam11}
{Simionescu} A.,  et~al., 2011, \mn@doi [Science] {10.1126/science.1200331},
  331, 1576

\bibitem[\protect\citeauthoryear{{Skillman}, {O'Shea}, {Hallman}, {Burns}  \&
  {Norman}}{{Skillman} et~al.}{2008}]{soh08}
{Skillman} S.~W.,  {O'Shea} B.~W.,  {Hallman} E.~J.,  {Burns} J.~O.,   {Norman}
  M.~L.,  2008, \mn@doi [\apj] {10.1086/592496}, 689, 1063

\bibitem[\protect\citeauthoryear{{Skillman}, {Xu}, {Hallman}, {O'Shea},
  {Burns}, {Li}, {Collins}  \& {Norman}}{{Skillman} et~al.}{2013}]{sxh13}
{Skillman} S.~W.,  {Xu} H.,  {Hallman} E.~J.,  {O'Shea} B.~W.,  {Burns} J.~O.,
  {Li} H.,  {Collins} D.~C.,   {Norman} M.~L.,  2013, \mn@doi [\apj]
  {10.1088/0004-637X/765/1/21}, 765, 21

\bibitem[\protect\citeauthoryear{{Stroe}, {Sobral}, {R{\"o}ttgering}  \& {van
  Weeren}}{{Stroe} et~al.}{2014}]{ssr14}
{Stroe} A.,  {Sobral} D.,  {R{\"o}ttgering} H.~J.~A.,   {van Weeren} R.~J.,
  2014, \mn@doi [\mnras] {10.1093/mnras/stt2286}, 438, 1377

\bibitem[\protect\citeauthoryear{{Stroe} et~al.,}{{Stroe} et~al.}{2016}]{ssr16}
{Stroe} A.,  et~al., 2016, \mn@doi [\mnras] {10.1093/mnras/stv2472}, 455, 2402

\bibitem[\protect\citeauthoryear{{Tozzi} \& {Norman}}{{Tozzi} \&
  {Norman}}{2001}]{tn01}
{Tozzi} P.,  {Norman} C.,  2001, \mn@doi [\apj] {10.1086/318237}, 546, 63

\bibitem[\protect\citeauthoryear{{Turk}, {Smith}, {Oishi}, {Skory}, {Skillman},
  {Abel}  \& {Norman}}{{Turk} et~al.}{2011}]{tso11}
{Turk} M.~J.,  {Smith} B.~D.,  {Oishi} J.~S.,  {Skory} S.,  {Skillman} S.~W.,
  {Abel} T.,   {Norman} M.~L.,  2011, \mn@doi [\apjs]
  {10.1088/0067-0049/192/1/9}, 192, 9

\bibitem[\protect\citeauthoryear{{Valdarnini}}{{Valdarnini}}{2011}]{valda11}
{Valdarnini} R.,  2011, \mn@doi [\aap] {10.1051/0004-6361/201015340}, 526, A158

\bibitem[\protect\citeauthoryear{{Vazza}, {Brunetti}, {Kritsuk}, {Wagner},
  {Gheller}  \& {Norman}}{{Vazza} et~al.}{2009}]{vbk09}
{Vazza} F.,  {Brunetti} G.,  {Kritsuk} A.,  {Wagner} R.,  {Gheller} C.,
  {Norman} M.,  2009, \mn@doi [\aap] {10.1051/0004-6361/200912535}, 504, 33

\bibitem[\protect\citeauthoryear{{Vazza}, {Brunetti}, {Gheller}  \&
  {Brunino}}{{Vazza} et~al.}{2010}]{vbg10}
{Vazza} F.,  {Brunetti} G.,  {Gheller} C.,   {Brunino} R.,  2010, \mn@doi [New
  Astronomy] {10.1016/j.newast.2010.05.003}, 15, 695

\bibitem[\protect\citeauthoryear{{Vazza}, {Dolag}, {Ryu}, {Brunetti},
  {Gheller}, {Kang}  \& {Pfrommer}}{{Vazza} et~al.}{2011a}]{vdr11}
{Vazza} F.,  {Dolag} K.,  {Ryu} D.,  {Brunetti} G.,  {Gheller} C.,  {Kang} H.,
   {Pfrommer} C.,  2011a, \mn@doi [\mnras] {10.1111/j.1365-2966.2011.19546.x},
  418, 960

\bibitem[\protect\citeauthoryear{{Vazza}, {Brunetti}, {Gheller}, {Brunino}  \&
  {Br{\"u}ggen}}{{Vazza} et~al.}{2011b}]{vbg11}
{Vazza} F.,  {Brunetti} G.,  {Gheller} C.,  {Brunino} R.,   {Br{\"u}ggen} M.,
  2011b, \mn@doi [\aap] {10.1051/0004-6361/201016015}, 529, A17

\bibitem[\protect\citeauthoryear{{Vazza}, {Roediger}  \& {Br{\"u}ggen}}{{Vazza}
  et~al.}{2012}]{vrb12}
{Vazza} F.,  {Roediger} E.,   {Br{\"u}ggen} M.,  2012, \mn@doi [\aap]
  {10.1051/0004-6361/201118688}, 544, A103

\bibitem[\protect\citeauthoryear{{Vazza}, {Eckert}, {Simionescu}, {Br{\"u}ggen}
   \& {Ettori}}{{Vazza} et~al.}{2013}]{ves13}
{Vazza} F.,  {Eckert} D.,  {Simionescu} A.,  {Br{\"u}ggen} M.,   {Ettori} S.,
  2013, \mn@doi [\mnras] {10.1093/mnras/sts375}, 429, 799

\bibitem[\protect\citeauthoryear{{Vazza}, {Gheller}  \& {Br{\"u}ggen}}{{Vazza}
  et~al.}{2014a}]{vgb14}
{Vazza} F.,  {Gheller} C.,   {Br{\"u}ggen} M.,  2014a, \mn@doi [\mnras]
  {10.1093/mnras/stu126}, 439, 2662

\bibitem[\protect\citeauthoryear{{Vazza}, {Br{\"u}ggen}, {Gheller}  \&
  {Wang}}{{Vazza} et~al.}{2014b}]{vbg14}
{Vazza} F.,  {Br{\"u}ggen} M.,  {Gheller} C.,   {Wang} P.,  2014b, \mn@doi
  [\mnras] {10.1093/mnras/stu1896}, 445, 3706

\bibitem[\protect\citeauthoryear{{Vazza}, {Jones}, {Br{\"u}ggen}, {Brunetti},
  {Gheller}, {Porter}  \& {Ryu}}{{Vazza} et~al.}{2017}]{vjb16}
{Vazza} F.,  {Jones} T.~W.,  {Br{\"u}ggen} M.,  {Brunetti} G.,  {Gheller} C.,
  {Porter} D.,   {Ryu} D.,  2017, \mn@doi [\mnras] {10.1093/mnras/stw2351},
  464, 210

\bibitem[\protect\citeauthoryear{{Voit}, {Kay}  \& {Bryan}}{{Voit}
  et~al.}{2005}]{vkb05}
{Voit} G.~M.,  {Kay} S.~T.,   {Bryan} G.~L.,  2005, \mn@doi [\mnras]
  {10.1111/j.1365-2966.2005.09621.x}, 364, 909

\bibitem[\protect\citeauthoryear{{Walker}, {Fabian}, {Sanders}, {Simionescu}
  \& {Tawara}}{{Walker} et~al.}{2013}]{wfs13}
{Walker} S.~A.,  {Fabian} A.~C.,  {Sanders} J.~S.,  {Simionescu} A.,   {Tawara}
  Y.,  2013, \mn@doi [\mnras] {10.1093/mnras/stt497}, 432, 554

\bibitem[\protect\citeauthoryear{{White}, {Briel}  \& {Henry}}{{White}
  et~al.}{1993}]{wbh93}
{White} S.~D.~M.,  {Briel} U.~G.,   {Henry} J.~P.,  1993, \mnras, 261, L8

\bibitem[\protect\citeauthoryear{{Xu}, {Li}, {Collins}, {Li}  \& {Norman}}{{Xu}
  et~al.}{2010}]{xlc10}
{Xu} H.,  {Li} H.,  {Collins} D.~C.,  {Li} S.,   {Norman} M.~L.,  2010, \mn@doi
  [\apj] {10.1088/0004-637X/725/2/2152}, 725, 2152

\bibitem[\protect\citeauthoryear{{Xu}, {Li}, {Collins}, {Li}  \& {Norman}}{{Xu}
  et~al.}{2011}]{xlc11}
{Xu} H.,  {Li} H.,  {Collins} D.~C.,  {Li} S.,   {Norman} M.~L.,  2011, \mn@doi
  [\apj] {10.1088/0004-637X/739/2/77}, 739, 77

\bibitem[\protect\citeauthoryear{{Zhuravleva}, {Churazov}, {Kravtsov}  \&
  {Sunyaev}}{{Zhuravleva} et~al.}{2012}]{zck12}
{Zhuravleva} I.,  {Churazov} E.,  {Kravtsov} A.,   {Sunyaev} R.,  2012, \mn@doi
  [\mnras] {10.1111/j.1365-2966.2012.20844.x}, 422, 2712

\bibitem[\protect\citeauthoryear{{Zhuravleva}, {Churazov}, {Kravtsov}, {Lau},
  {Nagai}  \& {Sunyaev}}{{Zhuravleva} et~al.}{2013}]{zck13}
{Zhuravleva} I.,  {Churazov} E.,  {Kravtsov} A.,  {Lau} E.~T.,  {Nagai} D.,
  {Sunyaev} R.,  2013, \mn@doi [\mnras] {10.1093/mnras/sts275}, 428, 3274

\bibitem[\protect\citeauthoryear{{Zhuravleva} et~al.,}{{Zhuravleva}
  et~al.}{2014}]{zcs14}
{Zhuravleva} I.,  et~al., 2014, \mn@doi [Nature] {10.1038/nature13830}, 515, 85

\bibitem[\protect\citeauthoryear{{Zinger}, {Dekel}, {Birnboim}, {Kravtsov}  \&
  {Nagai}}{{Zinger} et~al.}{2016}]{zdb15}
{Zinger} E.,  {Dekel} A.,  {Birnboim} Y.,  {Kravtsov} A.,   {Nagai} D.,  2016,
  \mn@doi [\mnras] {10.1093/mnras/stw1283}, 461, 412

\bibitem[\protect\citeauthoryear{{van Weeren}, {R{\"o}ttgering}, {Intema},
  {Rudnick}, {Br{\"u}ggen}, {Hoeft}  \& {Oonk}}{{van Weeren}
  et~al.}{2012}]{wri12}
{van Weeren} R.~J.,  {R{\"o}ttgering} H.~J.~A.,  {Intema} H.~T.,  {Rudnick} L.,
   {Br{\"u}ggen} M.,  {Hoeft} M.,   {Oonk} J.~B.~R.,  2012, \mn@doi [\aap]
  {10.1051/0004-6361/201219000}, \href
  {http://adsabs.harvard.edu/abs/2012A%26A...546A.124V} {546, A124}

\makeatother
\end{thebibliography}
\bibliographystyle{mnras}

\bsp
\label{lastpage}
\end{document}